\documentclass[a4paper,12pt]{article}
\pdfoutput=1

\usepackage[utf8x]{inputenc}
\usepackage{amsmath}
  \numberwithin{equation}{section}
   \allowdisplaybreaks
\usepackage{graphicx}
\usepackage{fullpage}
\usepackage[hyperfootnotes=false]{hyperref}

\usepackage{setspace}
\title{ \vspace*{\fill} \bf A simpler prescription for MHV graviton tree amplitudes in superstring theory}
\vspace{1cm}
\author{Tiago Ursulino\\{\normalsize ursulino@ift.unesp.br}\\~\\ \emph{Instituto de F\'{i}sica Te\'{o}rica, Universidade Estadual Paulista (IFT-Unesp)}\\ \emph{S\~{a}o Paulo, SP, Brazil}}
\date{\small September 2013} 

\begin{document}

\maketitle
\thispagestyle{empty}

\begin{abstract}

\indent We extend the Berkovits-Maldacena prescription for MHV amplitudes of the open superstring
to the closed superstring, showing that in the $\alpha'=0$ limit
it reduces to the result of supergravity found recently by Hodges.
We also verify that this prescription calculates the correct superstring tree level
MHV amplitude for 4 gravitons including $\alpha'$ corrections.

\end{abstract}


\vspace*{\fill}

\newpage
\setcounter{page}{1}

\tableofcontents

\section{Introduction}

\indent Berkovits and Maldacena \cite{berkovits} have proposed 5 years ago a prescription to calculate MHV amplitudes for gluons at tree level
inside the context of string theory, in a 4 dimensional spacetime. Their prescription (formula \eqref{gluons_prescription}) looks like that of an open superstring theory, and
they have shown that it gives the same result as open superstrings for amplitudes with $n=4$ and $5$ gluons including the massive
corrections ($\alpha'$ corrections) (Of course, as superstrings
live in 10 dimensions, one has to attach the open strings to D3-branes so that they can only move in the usual 4 dimensions.)
They have also shown that, in the limit $\alpha' = 0$, their prescription reduces to the formula proposed by Parke and Taylor \cite{parke,berends,nair} for Yang-Mills MHV amplitudes, as it should. Later, Stieberger and Taylor \cite{stieberger}
have also verified that it gives the right superstring result for $n=6$ gluons.\\
\indent Berkovits-Maldacena prescription is different from usual superstrings techniques in that: a) three of the vertex operators appearing in the expression
have only the exponential ``tachyonic'' factor and b) there are only ``physical'' fields appearing in all vertex operators, i.e., no ghosts and no pure spinors.\\
\indent And yet, it is not known how to relate it to an actual string theory. In other words, there is no known string action functional that
gives rise to such a prescription.\\
\indent The purpose of this work is to present the closed string analogue of the Berkovits-Maldacena prescription, and show that it
gives, for $n=4$ gravitons, the same result as with usual closed superstrings calculations, including massive corrections.\footnote{This was
earlier shown by Gustavo Monteiro in an unpublished work \cite{gustavo2}.} We are thinking here about closed superstrings with the extra 6 spatial dimensions compactified
and very small. As an original contribution, we show also that it reduces
to the known formulas for the MHV graviton amplitudes in the limit $\alpha'=0$ (see \cite{berends2,witten0,tranca,bedford,elvang,mason,nguyen,hodges2,hodges}). For this purpose we
will use the formula found by Hodges in \cite{hodges}. Again, the prescription only calculates
amplitudes at the tree level of string theory.\\
\indent Such a result might have interesting connections with $\mathcal{N}$ $=2$ closed string, which describes self-dual gravitons \cite{ooguri}, and it would also be interesting to extend it to
non-MHV amplitudes, as it has been done for supergravity recently (\cite{cachazo} and \cite{cachazo2}).


\section*{Notations}

\noindent We will be mostly interested in a 4 dimensional spacetime, so indices like $\mu$, $\nu$ will run from 0 to 3.\\
\indent Indices $\alpha$ and $\dot{\alpha}$ (and similars) are applied to $\mathbf{2}$ and $\mathbf{\bar{2}}$ spinor representations
of Lorentz 4 dimensional group.\\
\indent Wherever needed, the signature will be $\eta_{\mu \nu} = \mathrm{diag}(-+++)$. The normalization for the antisymmetric symbol $\epsilon_{\alpha \beta}$ (which is used to raise and lower spinor indices) will be $\epsilon_{21}=i\sqrt{2}$ and $\epsilon^{12}=i/\sqrt{2}$ (similarly for $\epsilon_{\dot{\alpha}\dot{\beta}}$), so that relation \eqref{metric} holds with no prefactor.\\
\indent The explicit forms of the Pauli matrices $\sigma^{\mu}_{\alpha \dot{\alpha}}$ are

\begin{align*}
&\sigma^0 = \left(
\begin{array}{cc}
 -1 & 0 \\
 0 & -1 \\
\end{array} \right) &\sigma^1 = \left(
\begin{array}{cc}
 0 & 1 \\
 1 & 0 \\
\end{array} \right)\\
&\sigma^2 = \left(
\begin{array}{cc}
 0 & -i \\
 i & 0 \\
\end{array} \right) &\sigma^3 = \left(
\begin{array}{cc}
 1 & 0 \\
 0 & -1 \\
\end{array} \right)
\end{align*}

\indent For the polarization vectors and spinors of gluons we will use the symbol $\varepsilon$ repeated times with different meanings. The
number and species of indices should be enough to differentiate the uses: $\varepsilon_{\mu}$ for the vector, $\varepsilon_{\alpha \dot{\alpha}}$ for the
corresponding $(\mathbf{2},\bar{\mathbf{2}})$ spinor and $\varepsilon_{\alpha}$ or $\varepsilon_{\dot{\alpha}}$ for specific choices of the polarization spinor
$\varepsilon_{\alpha \dot{\alpha}}$ as in expressions \eqref{polarizations}.\\
\indent Similarly the letter $h$ will be used with different meanings, all related to the polarization of the graviton.


\section{MHV amplitudes in super-Yang-Mills and supergravity}

\noindent This section makes a brief review of the subject. The experienced reader
may skip it and go to section \eqref{mhvsuperstring}.

\subsection{Helicity spinors}

\noindent All formulas of interest here will be written in terms of spinors instead of spacetime vectors.
Let us review how the former come up from the latter.\\
\indent Consider a massless particle with momentum $k^{\mu}$. Using the Pauli matrices $\sigma^{\mu}_{\alpha \dot{\alpha}}$ ($\alpha,\dot{\alpha}=1,2$), 
we can write the momentum as a matrix:

\begin{equation}
 k_{\alpha \dot{\alpha}} \equiv k_{\mu}\sigma^{\mu}_{\alpha \dot{\alpha}}
 \label{momentummatrix}
\end{equation}

\indent It is easy to see that $k^2 = -\det ( k_{\alpha \dot{\alpha}} )$ so that the ``matrix momentum'' of the particle
has null determinant. This means that this matrix can be factorized:

\begin{equation}
  k_{\alpha \dot{\alpha}} = \lambda_{\alpha} \tilde{\lambda}_{\dot{\alpha}} 
  \label{momentumasspinors}
\end{equation}

\noindent where $\lambda_{\alpha}$ and $\tilde{\lambda}_{\dot{\alpha}}$ are 2-component commuting spinors. We can verify that both sides of the above equation have the same number of independent components, but first let us discuss the Minkowskian
signature in use.\\
\indent With signature $\eta_{\mu \nu} = \mathrm{diag}(-+++)$ and using the explicit forms of the Pauli matrices, we have:

\begin{equation}
 k_{\alpha \dot{\alpha}} = \left(
\begin{array}{cc}
 -k_0+k_3 & k_1-i k_2 \\
 k_1+i k_2 & -k_0-k_3 \\
\end{array} \right)
= \left(
\begin{array}{cc}
 \lambda_1 \tilde{\lambda}_{\dot{1}} & \lambda_1 \tilde{\lambda}_{\dot{2}} \\
 \lambda_2 \tilde{\lambda}_{\dot{1}} & \lambda_2 \tilde{\lambda}_{\dot{2}} \\
\end{array} \right)
\label{kaslambda}
\end{equation}

\indent With this signature, hence, the spinors components must be complex and we have to set $\tilde{\lambda}_{\dot{\alpha}} =  \pm \bar{\lambda}_{\alpha}$ depending
if the particle is traveling to the past or to the future. If we perform Wick
rotations on some of the momentum components, these conditions change (e.g. if we Wick-rotate $k_2$, all 
matrix components will be real, and then we can also let the spinors components be real, with no relations between 
$\tilde{\lambda}_{\dot{\alpha}}$ and $\lambda_{\alpha}$).\\
\indent In any case, it looks like we have a total of 4 independent spinors components, while only 3 independent momentum matrix components
 because of the zero determinant condition. But look again at eq. \eqref{momentumasspinors}: this decomposition 
 will be the same if we rescale the spinors
 
 \begin{align}
 \lambda_{\alpha} &\rightarrow u \lambda_{\alpha} \nonumber \\
 \tilde{\lambda}_{\dot{\alpha}} &\rightarrow u^{-1} \tilde{\lambda}_{\dot{\alpha}} \label{rescalingspinors}
 \end{align}

\noindent where $u$ is a complex number of modulus 1 in the $(-+++)$ signature or a real number with $k_2$ Wick-rotated, and thus 
it takes away one component of the spinors, leaving 3 independent components on both sides of eq. \eqref{momentumasspinors}.\\
\indent Before moving on, let us note that we can also write the Minkowski tensor $\eta_{\mu \nu}$ as

\begin{equation}
  \eta_{\alpha \dot{\alpha}\beta \dot{\beta}} \equiv \eta_{\mu \nu} \sigma^{\mu}_{\alpha \dot{\alpha}}\sigma^{\nu}_{\beta \dot{\beta}} = \epsilon_{\alpha \beta}\epsilon_{\dot{\alpha} \dot{\beta}}
  \label{metric}
\end{equation}

\noindent where the last equality can be easily verified using the explicit forms of the Pauli matrices.\\
\indent Assume for an instant that the particle is a gluon (i.e. a spin 1 particle). Then it has a polarization vector $\varepsilon_{\mu}$, orthogonal to the particle momentum, $\varepsilon_{\mu}k^{\mu}=0$. We can
write the polarization vector as a matrix $\varepsilon_{\alpha \dot{\alpha}}$ like we did before for the momentum, and considering eq. \eqref{metric}
one can see that the condition of orthogonality is obeyed by the polarization vectors

\begin{equation}
\varepsilon^{+}_{\alpha \dot{\alpha}} \propto \varepsilon_{\alpha}\tilde{\lambda}_{\dot{\alpha}} \mathrm{,} \quad \varepsilon^{-}_{\alpha \dot{\alpha}} \propto \lambda_{\alpha}\tilde{\varepsilon}_{\dot{\alpha}}
\label{polarizations}
\end{equation}

\noindent for arbitrary normalization constants for $\varepsilon_{\alpha}$ or $\tilde{\varepsilon}_{\dot{\alpha}}$. The $+$ vector represents the polarization of a particle with positive helicity (angular momentum
pointing in the same direction of the momentum), and the $-$ that of a particle with negative helicity. The gauge transformation $\varepsilon_{\mu} \rightarrow \varepsilon_{\mu} + \Lambda k_{\mu}$ translates into

\begin{equation}
\varepsilon \rightarrow \varepsilon + \Lambda \lambda \quad \mathrm{or} \quad \tilde{\varepsilon} \rightarrow \tilde{\varepsilon} + \Lambda \tilde{\lambda}
\label{gaugeinvariance}
\end{equation}

\indent Similarly, gravitons have a polarization \textit{tensor} $h_{\mu\nu}$, and in terms of spinors it can be written in the form

\begin{equation}
  h^{+}_{\alpha \dot{\alpha} \beta \dot{\beta}} = h_{\alpha \beta} \tilde{\lambda}_{\dot{\alpha}} \tilde{\lambda}_{\dot{\beta}}
\label{polarization_graviton}
\end{equation}

\noindent for the $+$ helicity (with normalization $h_{\alpha \beta} \lambda^{\alpha} \lambda^{\beta} = 1 $) and

\begin{equation}
 h^{-}_{\alpha \dot{\alpha} \beta \dot{\beta}} = \lambda_{\alpha} \lambda_{\beta} \tilde{h}_{\dot{\alpha} \dot{\beta}}
 \label{polarization_graviton2}
\end{equation}

\noindent for the $-$ helicity (with normalization $\tilde{\lambda}_{\dot{\alpha}} \tilde{\lambda}_{\dot{\beta}} \tilde{h}^{\dot{\alpha} \dot{\beta}} = 1 $).\footnote{This normalization makes $h^+_{\mu\nu}h^{-\mu\nu}=1$.} Of course,
because of the symmetry of $h_{\mu\nu}$, $h_{\alpha\beta}$ and $\tilde{h}_{\dot{\alpha}\dot{\beta}}$ are also symmetric.\\
\indent The gauge transformations for the gravitons are:

\begin{align}
  & \delta h_{\alpha \beta} = \zeta_{\alpha} \lambda_{\beta}+\zeta_{\beta} \lambda_{\alpha} \nonumber \\
  & \delta \tilde{h}_{\dot{\alpha} \dot{\beta}} = \tilde{\zeta}_{\dot{\alpha}} \tilde{\lambda}_{\dot{\beta}}+\tilde{\zeta}_{\dot{\beta}} \tilde{\lambda}_{\dot{\alpha}} \label{gauge_graviton}
\end{align}

\noindent where $\zeta_{\alpha}$, $\tilde{\zeta}_{\dot{\alpha}}$ are arbitrary spinors.\\
\indent Let us also define the invariant spinor products:

\begin{align}
  &( \lambda \xi ) \equiv  \lambda^{\alpha} \xi_{\alpha}= \epsilon_{\alpha \beta} \lambda^{\alpha} \xi^{\beta} = - \xi^{\beta}\lambda_{\beta}=-( \xi \lambda ) \nonumber \\
  &( \tilde{\lambda} \tilde{\xi} ) \equiv \tilde{\lambda}^{\dot{\alpha}} \tilde{\xi}_{\dot{\alpha}} = -( \tilde{\xi} \tilde{\lambda} ) \label{spinor_prod}
\end{align}

\indent In particular:

\[ (\lambda\lambda)=(\tilde{\lambda}\tilde{\lambda})=0 \]

\indent One property of interest in our calculations is the following: consider a scattering between
a set of $n$ incoming and outgoing (massless) particles of momenta $k_i = \lambda_i \tilde{\lambda}_i$; by total momentum conservation, we have:

\[ \sum_{i \text{ is an incoming particle}} (\lambda_i)_{\alpha}(\tilde{\lambda}_i)_{\dot{\alpha}} = \sum_{j \text{ is an outgoing particle}} (\lambda_j)_{\alpha}(\tilde{\lambda}_j)_{\dot{\alpha}} \]

\indent We can change the point of view by passing the RHS of the equation above to the LHS and treating the
outgoing particles as incoming, but with reversed momenta, absorbing the minus sign with the $\tilde{\lambda}_i$'s in
$\eta = \text{diag}(-+++)$ signature, for instance. In other words, we can think of this scattering as
a process involving future and past traveling particles with vanishing total momentum:

\[ \sum_{i=1}^n (\lambda_i)_{\alpha}(\tilde{\lambda}_i)_{\dot{\alpha}} = 0\]

\indent By contracting the expression above in the left and the right with arbitrary $\lambda_r$ and $\tilde{\lambda}_s$, we have:

\begin{equation}
\sum_{i=1}^n (\lambda_r\lambda_i)(\tilde{\lambda}_i\tilde{\lambda}_s)=0
 \label{conservationproperty}
\end{equation}

\indent In particular, if $r=s$ and defining $s_{ij} \equiv k_i \cdot k_j$, we have

\begin{equation}
\sum_{i=1}^n s_{ri} = 0
 \label{conservationproperty2}
\end{equation}

\indent This is the point of view we will use throughout this work.

\subsection{Supersymmetric extensions of YM and gravity}

\noindent The formulas for MHV amplitudes we will be concerned with here are valid not only to YM or gravity, but also to
their extensions with maximum number of supersymmetries.\\
\indent If $Q^i_{\alpha}$ and $\bar{Q}^j_{\dot{\alpha}}$ are the generators of the supersymmetries, we have

\begin{align}
\{Q^i_{\alpha},\bar{Q}^j_{\dot{\alpha}}\} &= 2\sigma^{\mu}_{\alpha \dot{\alpha}}P_{\mu}\delta^{ij}\\
\{Q^i_{\alpha},Q^j_{\alpha}\} &= \{\bar{Q}^i_{\dot{\alpha}},\bar{Q}^j_{\dot{\alpha}}\} = 0 \nonumber 
 \label{anticommutationofqs}
\end{align}

\noindent where $P_{\mu}$ is the generator of spacetime translations. The indices $i,j$ run from $1$ to an $N$ defining the number of symmetries of the theory. These supersymmetries are
usually better visualized as translations over a bigger spacetime, described not only by the coordinates $x^{\mu}$ but also by Grassmann variables $\theta^{\alpha i}$ and $\bar{\theta}_{\dot{\alpha} i}$.\\
\indent A YM theory can be extended to a sYM theory with up to $N=4$ supersymmetries. This $N=4$ sYM theory contains $2^4=16$ states: 
the gluon (with 2 states, the + and - helicity states), fermions of $1/2$ spin (8 states) and 6 scalars. We can join these states in a
single ``superstate'' with the help of four Grassmann variables $\kappa_i$ ($i=1,...,4$). Say $a^+(\lambda,\tilde{\lambda})$ represents the state of the positive helicity gluon; the full superstate will be written in the form:

\begin{equation}
   f(\lambda,\tilde{\lambda},\kappa_i) = a^+(\lambda,\tilde{\lambda})+\kappa_i\psi^i(\lambda,\tilde{\lambda}) + \kappa_i\kappa_j\phi^{ij}(\lambda,\tilde{\lambda}) +...+\kappa_1\kappa_2\kappa_3\kappa_4 a^-(\lambda,\tilde{\lambda})
 \label{superstate}
\end{equation}

\indent Now, while the wave function of a single gluon (of positive helicity, say) would be a function of $\lambda,\tilde{\lambda}$ times the exponential $e^{i\lambda^{\alpha}\tilde{\lambda}^{\dot{\alpha}}x_{\alpha \dot{\alpha}}}$, we can think
about a ``super wave function'', with the variables $\kappa_i$ conjugated to the variables $\lambda_{\alpha}\theta^{\alpha i}$:

\begin{equation}
 f(\lambda,\tilde{\lambda},\kappa_i)e^{i\lambda^{\alpha}\tilde{\lambda}^{\dot{\alpha}}x_{\alpha \dot{\alpha}}}e^{\lambda_{\alpha}\kappa_i\theta^{\alpha i}}
\label{superwave}
\end{equation}

\indent Expanding the exponential in powers of the $\kappa_i$ we recover the various components of the super wave function \cite{berkovits}. In all this discussion 
we are forgetting about the color indices, but of course all the particles in the supermultiplet must be in the same representation as the gluon, i.e. the
adjoint representation of the gauge group.\\
\indent A scattering amplitude in the sYM theory will be a function of the momenta of the particles involved but also of these new parameters
$\kappa_i$. To obtain the amplitude of the scattering between, say, $n$ + helicity gluons, we would have to look at the lowest order term (in the $\kappa_i$'s) in
the expansion of the amplitude. More explicitly, the same way the integral over spacetime will produce a total momentum preserving delta function $\delta^4(\sum_r k_r)$,
an integral over the variables $\theta$ will produce a Grassmannian delta function, and the amplitude can be written in the form

\begin{equation}
 A_n(\lambda^r,\tilde{\lambda}^r,\kappa^r_i) = i(2\pi)^4\delta^4 \left( \sum_{r=1}^n k_r^{\mu}\right) \delta^8 \left( \sum_{r=1}^n \lambda_{\alpha}^r\kappa^r_i \right)\tilde{A}_n(\lambda^r,\tilde{\lambda}^r,\kappa^r_i)
 \label{decomposition}
\end{equation}

\indent (This is the color ordered amplitude, with the color indices of the gauge group omitted.)\\
\indent Remember that the delta function of a Grassmann variable $\xi$ is just $\delta(\xi)=\xi$. Now we can actually see that, expanding the amplitude in powers of $\kappa$'s, the term in $(\kappa_i^r)^0$ for all $r$'s and $i$'s vanishes; this
corresponds to the scattering amplitude between $n$ gluons of $+$ helicity. The amplitude with just one - helicity gluon (say, the gluon $s$) corresponds to the term
in $\kappa^s_1\kappa^s_2\kappa^s_3\kappa^s_4$, with all other $\kappa$'s to the 0-th power; this term also vanishes. The first possibly non-vanishing terms are the ones
with $8$ $\kappa$'s, or, in other words, they are the lowest orders terms of the tilded amplitude $\tilde{A}_n$: these are the amplitudes we
are going to call MHV (maximally helicity violating) from now on.\\
\indent We see, hence, that to extract the amplitude of the scattering where only gluons $s$ and $t$ have - helicity we have to
perform the Berezin integrals over the corresponding $\kappa$ variables of those gluons, i.e., integrate over $d\kappa^s_1 d\kappa^s_2 d\kappa^s_3 d\kappa^s_4 d\kappa^t_1 d\kappa^t_2 d\kappa^t_3 d\kappa^t_4$; the factors
of $\lambda^s$ and $\lambda^t$ are left to be combined in a $SU(2)$ invariant form; we know this form: it is the product $(\lambda_s \lambda_t)$ defined in the last section. There are
four of these products; the MHV amplitude can be written, thus, as

\begin{equation}
 A_n^{MHV}(\lambda^r,\tilde{\lambda}^r) = i(2\pi)^4\delta^4 \left( \sum_{r=1}^n k_r^{\mu}\right) (\lambda_s \lambda_t)^4 \tilde{A}_n(\lambda^r,\tilde{\lambda}^r,0)
 \label{factorgluons}
\end{equation}

\indent One can also talk about MHV amplitudes between gluons and gluinos by performing different combinations of Berezin integrals over the $\kappa$ variables.\\
\indent In the next section we will see the surprisingly simple formula Parke and Taylor \cite{parke} proposed for the tilded amplitude $\tilde{A}_n(\lambda^r,\tilde{\lambda}^r,0)$.\footnote{In their article, actually, they were not talking about sYM, so they
have included the prefactor of $(\lambda_s \lambda_t)^4$ in the formula \eqref{mhv_gluon}. The treatment in superspace was first done by Nair \cite{nair}. Also, Parke and Taylor presented the formula for the squared amplitude, by means of which they did not need to use the language of spinors.}\\
\indent The case of supergravity with maximum number of supersymmetries ($N=8$ supergravity) is treated similarly: one simply needs to make the indices
$i$ run from $1$ to $8$ now. Integrals over the $\theta^{\alpha i}$ variables now will produce a $16$ dimensional Grassmann delta function for the $\kappa$'s. If we are considering
the amplitude where only gravitons $s$ and $t$ have - helicity, then as in the expression above we have:

\begin{equation}
 M_n^{MHV}(\lambda^r,\tilde{\lambda}^r) = i(2\pi)^4\delta^4 \left( \sum_{r=1}^n k_r^{\mu}\right) (\lambda_s \lambda_t)^8 \tilde{M}_n(\lambda^r,\tilde{\lambda}^r,0)
 \label{factorgravitons}
\end{equation}

\indent We will explore in detail the formulas proposed for the tilded amplitude $\tilde{M}_n(\lambda^r,\tilde{\lambda}^r,0)$ for the case of gravitons.\\
\indent Again, we can also talk about more general MHV gravitational amplitudes, involving gravitinos and other particles in the supermultiplet. Throughout this work
we will mostly talk about gravitons (or gluons), but one should keep in mind that the formulas proposed can be used in those more general MHV configurations.\\
\indent Let us finally notice that, in formula \eqref{decomposition}, the ``decision'' about which particles will carry - helicity
had not been taken yet: in that formula the particles enter in a completely symmetric way,\footnote{For the gluons, there is only a cyclic symmetry, as already discussed, but for the gravitons
there is a full $n$ objects permutation symmetry} and only via the Berezin integrals
we ``select'' the states involved in the amplitude we are interested in. That means, also, that the tilded amplitudes are symmetric
between the different particles, property that it will be important to keep in mind in the next sections.

\subsection{MHV amplitudes for gluons}

\indent MHV amplitudes are interesting because they are much simpler to calculate than general gluon scattering amplitudes, having an
expression that enables one to skip the Feynman diagrams method. The formula proposed by Parke and Taylor \cite{parke}, proved by Berends and Giele \cite{berends} and extended to sYM by Nair \cite{nair}, and in the notation of the last section, is:

\begin{equation}
  \tilde{A}_n^{MHV} = g^{n-2} \frac{1}{\prod_{r=1}^{n} (\lambda_r \lambda_{r+1})}
\label{mhv_gluon}
\end{equation}

\noindent where $g$ is the coupling constant of the theory and $\lambda_{n+1} \equiv \lambda_1$. Here, the superscript $MHV$ means, of course, that
the parameters $\kappa$ of the last section are taken to be $0$, but we will stop writing it in the future.\\
\indent Again, this is the \textit{color ordered} amplitude, i.e., the amplitude
for the scattering of the gluons with the prefactor of the traces over the gauge group generators removed.\\
\indent We can see immediately that this expression has a cyclic symmetry, as we anticipated it should have.\\
\indent The formula was first proved correct in \cite{berends} by the use of recursion relations between amplitudes with
different numbers of gluons. (See also \cite{gustavo}.)

\subsection{MHV amplitudes for gravitons}

\noindent MHV amplitudes for gravitons are also simpler to calculate, though far more involved than the corresponding ones for gluons. First we will present
and discuss here some of the formulas available, using later two of them to calculate
MHV amplitudes for 4 and 5 gravitons.

\subsubsection{An overview of the formulas so far; symmetry and efficiency}

\indent It has been almost 25 years since the first formula for such amplitudes was found by Berends, Giele and Kuijf \cite{berends2}. They have
used string theory to obtain their result, exploring the relation between closed and open strings (results for closed strings
being somewhat the ``square'' of the results for open strings \cite{klt}) and the result for gluons eq. \eqref{mhv_gluon}. Their result was:

\begin{equation}
  \tilde{M}_n^{MHV} = \left( \kappa \right)^{n-2} \frac{(\tilde{\lambda}_1\tilde{\lambda}_2)(\tilde{\lambda}_{n-2}\tilde{\lambda}_{n-1})}{(\lambda_1 \lambda_{n-1})}\frac{F}{N(n)}\prod_{i=3}^{n-3}\prod_{j=i+2}^{n-1}(\lambda_i\lambda_j) + P_{(2,...,n-2)}
 \label{mhvberends}
\end{equation}

\noindent where $\kappa$ stands for the gravitational coupling constant, $P_{(2,...,n-2)}$ for the sum over all permutations between the indices indicated,

\[ N(n) \equiv \prod_{i<j} (\lambda_i \lambda_j) \]

\noindent and

\[ F \equiv \prod_{k=3}^{n-3}\{ (\tilde{\lambda}_k \tilde{\lambda}_{k+1})(\lambda_{k+1}\lambda_n) + (\tilde{\lambda}_k \tilde{\lambda}_{k+2})(\lambda_{k+2}\lambda_n) + ... + (\tilde{\lambda}_k \tilde{\lambda}_{n-1})(\lambda_{n-1}\lambda_n) \} \]

\indent It is not important for us to understand the details of this formula, as it will not be used to be
compared with our results in the next chapter. Let us just make some comments.\\
\indent First of all, there doesn't seem to be a full $S_n$ symmetry\footnote{$S_n$ standing for the group of permutations of $n$ objects.} between the gravitons
involved, but only a $S_{n-3}$ symmetry, and that too only because of the explicit presence of the sum over the many permutations. Of course, the full symmetry is there,
only hidden.\\
\indent The presence of the sum over $S_{n-3}$ permutations also makes this formula very slow from
the point of view of number of operations involved in the calculation: there are $(n-3)!$ such permutations, so the number of operations is at least of that order.\\
\indent Mason and Skinner \cite{mason} simplified the BKG formula a little:

\begin{equation}
\tilde{M}_n^{MHV} = \left( \kappa \right)^{n-2} \frac{1}{(\lambda_1 \lambda_{n-1})(\lambda_{n-1} \lambda_n)(\lambda_n \lambda_1)}\frac{F'}{C(n)}+P_{(2,...,n-2)}
 \label{mhvmason}
\end{equation}

\noindent where $C(n)$ is a cyclic product identical to the one inside the gluons amplitude, eq. \eqref{mhv_gluon},

\[ C(n) = \prod_{i=1}^n(\lambda_i\lambda_{i+1}) \]

\noindent and $F'$, similar to the $F$ above, is

\[ F' \equiv \prod_{k=2}^{n-1} \frac{ (\tilde{\lambda}_k \tilde{\lambda}_{k+1})(\lambda_{k+1}\lambda_n) + (\tilde{\lambda}_k \tilde{\lambda}_{k+2})(\lambda_{k+2}\lambda_n) + ... + (\tilde{\lambda}_k \tilde{\lambda}_{n-1})(\lambda_{n-1}\lambda_n) }{(\lambda_k \lambda_n)} \]

\indent This formula, obtained using twistor formalism, apart from the obvious simplification over BKG formula, does not bring much
more advantages: it still has an explicit sum over permutations, which guarantees only a partial $S_{n-3}$ symmetry. The simplicity
also quickens a little the calculation, but the number of operations still remains at least of the order of $(n-3)!$.\\
\indent Another formula was proposed by Bedford et al. \cite{bedford} which contains a sum over permutations of $(n-2)$
of the indices. This improves the symmetry from $S_{n-3}$ to $S_{n-2}$, not improving the calculational efficiency though.\\
\indent An interesting formula was proposed by Elvang and Freedman \cite{elvang}, in which is clear the expression of the gravitational
amplitude as a sum over squares of YM amplitudes:

\begin{equation}
\tilde{M}_n(1,2,3,...,n)= (\lambda_1 \lambda_n)(\tilde{\lambda}_1\tilde{\lambda}_n)\left( \prod_{i=4}^{n-1} \beta_i \right) \tilde{A}_n(1,2,3,...,n)^2+P_{(3,4,...,n)}
 \label{mhvelvang}
\end{equation}

\noindent where

\[ \beta_i \equiv -\frac{(\lambda_i \lambda_{i+1})}{(\lambda_2 \lambda_{i+1})}\{(\lambda_2\lambda_3)(\tilde{\lambda}_3\tilde{\lambda}_i)+(\lambda_2\lambda_4)(\tilde{\lambda}_4\tilde{\lambda}_i)+...+(\lambda_2\lambda_{i-1})(\tilde{\lambda}_{i-1}\tilde{\lambda}_i)\} \]

\noindent and the tilded $\tilde{A}_n$ amplitude is the one for gluons as in formula \eqref{mhv_gluon}.\\
\indent Again there is an explicit summation over permutations, but here over $(n-2)$ indices. The
coupling constants are taken here to be $1$, or one may think that they have been factored out of the tilded
amplitudes in eq. \eqref{decomposition}.\\
\indent One can check for the simplest cases the equivalence between all these formulas, or look in the articles
cited for the general proofs. We will check specific cases for the more interesting formulas given below.\\

\subsubsection*{Tree formula}

\noindent Nguyen et al. \cite{nguyen} have proposed an interesting method to calculate those amplitudes.
Their formula makes use of diagrams that they called ``trees'' (which are not regular tree Feynman diagrams, though). To write it one has
to choose 2 particles that will enter the expression in a special way (any 2 will do, there is no relation
between this choice and the two gravitons with different helicity, because this choice is not to be made at this point, as we have
already discussed). Of course, this breaks the $S_n$ permutation symmetry to only an $S_{n-2}$ symmetry again. Here is the formula, if
we choose to treat gravitons $(n-1)$ and $n$ specially:

\begin{equation}
\tilde{M}_n = \frac{1}{(\lambda_{n-1}\lambda_n)^2}\sum_{\text{trees}}\left( \prod_{\text{edges }ab} \frac{(\tilde{\lambda}_a\tilde{\lambda}_b)}{(\lambda_a \lambda_b)} \right) \left( \prod_{\text{vertices} a} [(\lambda_a\lambda_{n-1})(\lambda_a\lambda_{n})]^{\deg(a)-2} \right)
 \label{mhvtreeformula}
\end{equation}

\indent And here is the meaning of this formula:

\begin{itemize}
 \item Choose two particles to have special treatment ($(n-1)$ and $n$ for instance) and put the overall factor $(\lambda_{n-1}\lambda_n)^{-2}$,
 \item With all other particles, draw a tree diagram, putting the label of the particles in the vertices and joining them with edges;
 \item For each edge connecting particles $a$ and $b$, put a factor of $\frac{(\tilde{\lambda}_a\tilde{\lambda}_b)}{(\lambda_a \lambda_b)}$;
 \item Each vertex $a$ has a degree $\deg(a)$, which is defined to be the number of edges connected to it;
 \item For each vertex, add the factor $[(\lambda_a\lambda_{n-1})(\lambda_a\lambda_{n})]^{\deg(a)-2}$, which ``contracts'' the particle in the vertex
 with the two particles being treated specially;
 \item Do the same for all possible distinct tree diagrams drawable.
\end{itemize}

\indent This method brings a small upgrade to the number of operations necessary, because there are $(n-2)^{(n-4)}$ diagrams for an $n$ gravitons amplitude.\\
\indent The case $n=4$ is almost trivial: treating $3$ and $4$ specially, there is only one diagram (figure \ref{n4diagram}). This diagram corresponds to the following expression:

\begin{figure}[h]
 \vspace{1cm}
\centering \includegraphics[width=0.25\textwidth]{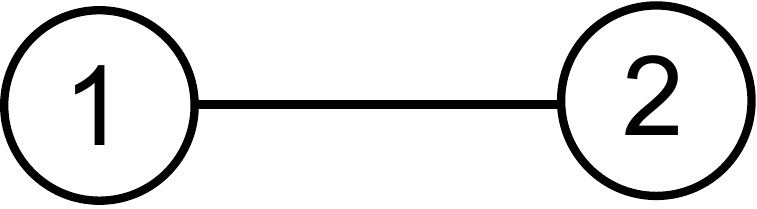}
 \caption{Tree diagram for $n=4$ gravitons MHV amplitude.}
 \label{n4diagram}
\end{figure}

\begin{equation}
 \tilde{M}_4 = \frac{1}{(\lambda_3\lambda_4)^2}\frac{(\tilde{\lambda}_1\tilde{\lambda}_2)}{(\lambda_1 \lambda_2)}\frac{1}{(\lambda_1\lambda_3)(\lambda_1\lambda_4)}\frac{1}{(\lambda_2\lambda_3)(\lambda_2\lambda_4)}
 \label{n4diagram_expression}
\end{equation}

\indent For the case of $n=5$ gravitons, there are 3 diagrams (choosing $4$ and $5$ for special treatment); see figure \ref{n5diagram}.
They correspond to:

\begin{figure}[h]
 \vspace{1cm}
\centering \includegraphics[width=0.25\textwidth]{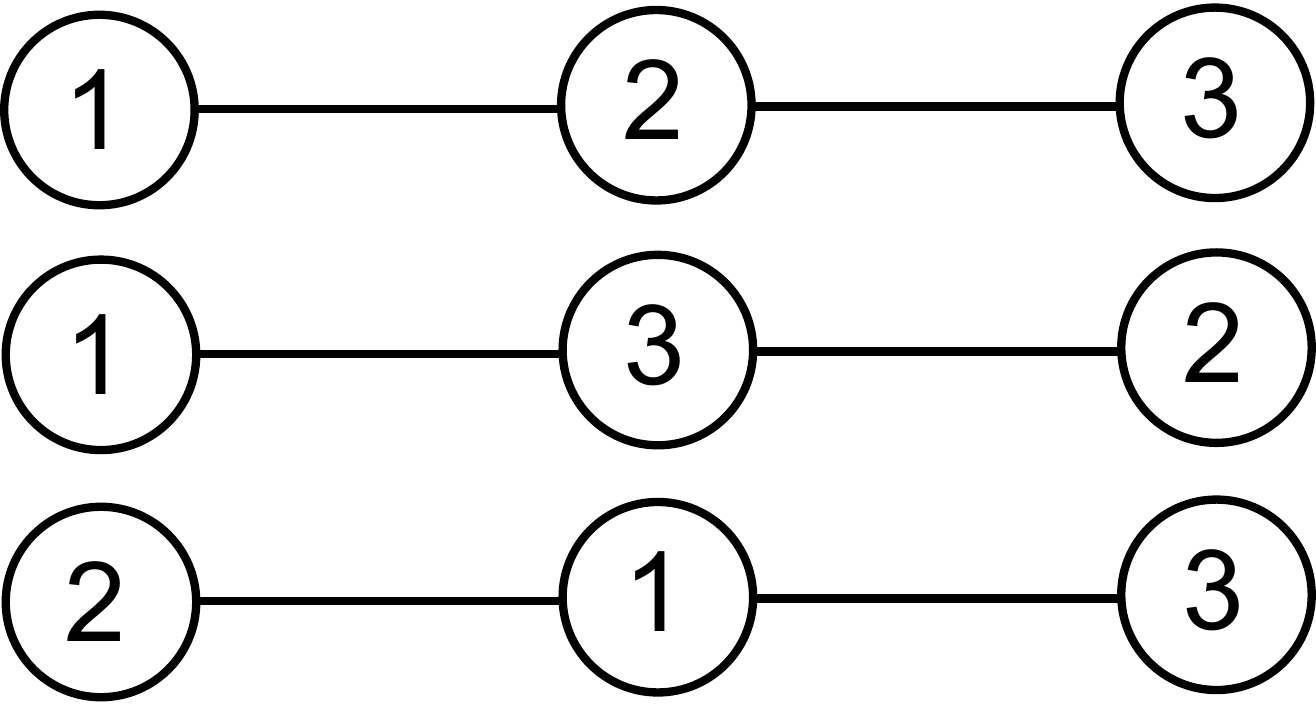}
 \caption{Tree diagrams for $n=5$ gravitons MHV amplitude.}
 \label{n5diagram}
\end{figure}

\begin{align}
 \tilde{M}_5 &= \frac{1}{(\lambda_4 \lambda_5)^2} \times \nonumber \\
 &\frac{(\tilde{\lambda}_1\tilde{\lambda}_2)}{(\lambda_1 \lambda_2)}\frac{(\tilde{\lambda}_2\tilde{\lambda}_3)}{(\lambda_2 \lambda_3)}\frac{1}{(\lambda_1\lambda_4)(\lambda_1\lambda_5)}\frac{1}{(\lambda_3\lambda_4)(\lambda_3\lambda_5)}\times  \label{n5diagram_expression}\\
 &\frac{(\tilde{\lambda}_1\tilde{\lambda}_3)}{(\lambda_1 \lambda_3)}\frac{(\tilde{\lambda}_3\tilde{\lambda}_2)}{(\lambda_3 \lambda_2)}\frac{1}{(\lambda_1\lambda_4)(\lambda_1\lambda_5)}\frac{1}{(\lambda_2\lambda_4)(\lambda_2\lambda_5)}\times \nonumber \\
 &\frac{(\tilde{\lambda}_2\tilde{\lambda}_1)}{(\lambda_2 \lambda_1)}\frac{(\tilde{\lambda}_1\tilde{\lambda}_3)}{(\lambda_1 \lambda_3)}\frac{1}{(\lambda_2\lambda_4)(\lambda_2\lambda_5)}\frac{1}{(\lambda_3\lambda_4)(\lambda_3\lambda_5)} \nonumber
\end{align}

\subsubsection*{Hodges' formula}

\indent The best formula so far is the one proposed by Hodges \cite{hodges} in 2012 (see also \cite{hodges2}):
it has the desired explicit $S_n$ symmetry \textit{without} making use of an explicit
sum over permutations. It also has a much quicker behavior in the number of operations.\\
\indent Here is the formula:

\begin{equation}
  \tilde{M}_n(1,2,...,n) = (-1)^{n+1}\text{sgn}({\alpha \beta}) c_{\alpha(1)\alpha(2)\alpha(3)}c^{\beta(1)\beta(2)\beta(3)}\phi^{\beta(4)}_{[\alpha(4)}\phi^{\beta(5)}_{\alpha(5)}...\phi^{\beta(n)}_{\alpha(n)]}
\label{mhvhodges}
\end{equation}

\noindent with the antisymmetrization taken with no combinatoric factors, and with the following definitions:

\begin{align}
  &\alpha\text{ and }\beta \text{ are arbitrary permutations of } (1,2,...,n) \nonumber \\
  &\text{sgn}(\alpha \beta)\text{ is the product of the signs of the permutations} \nonumber \\
  &c_{ijk}=c^{ijk}\equiv [ (\lambda_i \lambda_j)(\lambda_j \lambda_k)(\lambda_k \lambda_i) ]^{-1} \label{hodges_definitions}\\
  &\phi^i_j \equiv \frac{(\tilde{\lambda}_i \tilde{\lambda}_j)}{(\lambda_i \lambda_j)}\text{ for } i \neq j \nonumber \\
  &\phi^i_i \equiv - \sum_{j=1,j\neq i}^{n} \frac{(\tilde{\lambda}_i \tilde{\lambda}_j)(\lambda_j x_i)(\lambda_j y_i)}{(\lambda_i \lambda_j)(\lambda_i x_i)(\lambda_i y_i)} \text{, where } x_i \text{ and } y_i \text{ are arbitrary spinors.}\nonumber
\end{align}

\indent Note that we are not taking a \textit{sum} over the permutations $\alpha$ and $\beta$! As the formula is correct
for any choice of those permutations, the symmetry between the gravitons is explicit.\\
\indent One might argue that there is actually a sum over permutations between the indices of the $\phi$'s, but
note that this sum is just the determinant of a symmetric matrix, and there are simple numerical methods to calculate this.
Indeed, to calculate the determinant of a symmetric $(n-3)\times(n-3)$ matrix there are $O(n^3)$ operations.\\
\indent Let us compare it with our previous results for the cases $n=4$ and $n=5$ found using the tree diagrams.\\
\indent For $n=4$, we will choose $\alpha(1,2,3,4)=\beta(1,2,3,4)=(2,3,4,1)$; with this choice, $\text{sgn}(\alpha\beta)=1$. Then one must worry only about $\phi_1^1$; if we choose $x_1 = \lambda_3$ and $y_1=\lambda_4$, we have:

\[ \phi_1^1 = - \frac{(\tilde{\lambda}_1\tilde{\lambda}_2)(\lambda_2\lambda_3)(\lambda_2\lambda_4)}{(\lambda_1\lambda_2)(\lambda_1\lambda_3)(\lambda_1\lambda_4)} \]

\indent Then the result is

\begin{align}
\tilde{M}_4 &= (-1)\frac{1}{(\lambda_2\lambda_3)^2 (\lambda_3\lambda_4)^2 (\lambda_4\lambda_2)^2}\phi_1^1 \label{n4hodges}\\
 &= \frac{1}{(\lambda_3\lambda_4)^2}\frac{(\tilde{\lambda}_1\tilde{\lambda}_2)}{(\lambda_1 \lambda_2)}\frac{1}{(\lambda_1\lambda_3)(\lambda_1\lambda_4)}\frac{1}{(\lambda_2\lambda_3)(\lambda_2\lambda_4)} \nonumber
\end{align}

\noindent which is the same result as found before (formula \eqref{n4diagram_expression}).\\
\indent For the case $n=5$, we will choose $\alpha(1,2,3,4,5)=\beta(1,2,3,4,5)=(3,4,5,1,2)$ ($\text{sgn}(\alpha\beta)=1$), with $x_1=x_2=\lambda_4$ and $y_1=y_2=\lambda_5$. Then we have

\begin{align*}
  \phi^1_{[1}\phi^2_{2]} &= \left[ \frac{(\tilde{\lambda}_1\tilde{\lambda}_2)(\lambda_2\lambda_4)(\lambda_2\lambda_5)}{(\lambda_1\lambda_2)(\lambda_1\lambda_4)(\lambda_1\lambda_5)}+\frac{(\tilde{\lambda}_1\tilde{\lambda}_3)(\lambda_3\lambda_4)(\lambda_3\lambda_5)}{(\lambda_1\lambda_3)(\lambda_1\lambda_4)(\lambda_1\lambda_5)} \right]\times \\
  &\left[ \frac{(\tilde{\lambda}_2\tilde{\lambda}_1)(\lambda_1\lambda_4)(\lambda_1\lambda_5)}{(\lambda_2\lambda_1)(\lambda_2\lambda_4)(\lambda_2\lambda_5)}+\frac{(\tilde{\lambda}_2\tilde{\lambda}_3)(\lambda_3\lambda_4)(\lambda_3\lambda_5)}{(\lambda_2\lambda_3)(\lambda_2\lambda_4)(\lambda_2\lambda_5)} \right]-\\
  &-\frac{(\tilde{\lambda}_1\tilde{\lambda}_2)^2}{(\lambda_1\lambda_2)^2}=
\end{align*}
\begin{align*}
  &=\frac{(\tilde{\lambda}_1\tilde{\lambda}_2)}{(\lambda_1\lambda_2)}\frac{(\tilde{\lambda}_2\tilde{\lambda}_3)}{(\lambda_2\lambda_3)}\frac{1}{(\lambda_1\lambda_4)(\lambda_1\lambda_5)}(\lambda_3\lambda_4)(\lambda_3\lambda_5)+\\
  &+\frac{(\tilde{\lambda}_1\tilde{\lambda}_3)}{(\lambda_1\lambda_3)}\frac{(\tilde{\lambda}_3\tilde{\lambda}_2)}{(\lambda_3\lambda_2)}\frac{1}{(\lambda_1\lambda_4)(\lambda_1\lambda_5)}\frac{1}{(\lambda_2\lambda_4)(\lambda_2\lambda_5)}(\lambda_3\lambda_4)^2(\lambda_3\lambda_5)^2+\\
  &+\frac{(\tilde{\lambda}_2\tilde{\lambda}_1)}{(\lambda_2\lambda_1)}\frac{(\tilde{\lambda}_1\tilde{\lambda}_3)}{(\lambda_1\lambda_3)}\frac{1}{(\lambda_2\lambda_4)(\lambda_2\lambda_5)}(\lambda_3\lambda_4)(\lambda_3\lambda_5)
\end{align*}

\noindent which, with the prefactor of $c_{345}c^{345}$ reduces to the result of eq. \eqref{n5diagram_expression}.\\


\section{MHV amplitudes in superstring theory}

\label{mhvsuperstring}

\subsection{Berkovits-Maldacena prescription}

\noindent Berkovits and Maldacena have proposed \cite{berkovits} a prescription to calculate MHV amplitudes for gluons at tree level of string theory, i.e.,
amplitudes which include also the interaction of the gluons with the massive string states.\\
\indent Explicitly, the prescription looks like:

\begin{equation}
 A_n(\lambda_i, \tilde{\lambda}_i) = \frac{1}{(\lambda_1\lambda_2)(\lambda_2\lambda_3)(\lambda_3\lambda_1)} \left\langle V(y_1)V(y_2)V(y_3)\prod_{r=4}^{n}\int_{y_{r-1}}^{y_1}d y_r U(y_r) \right\rangle
\label{gluons_prescription}
\end{equation}
\noindent where the $y$'s are real variables,\footnote{Because we are talking about open strings,
and describing the world-sheet as the upper half-complex plane. The vertex
operators lie in the boundary, which is the real line in this description.}
$V(y)=\exp{(i\lambda^{\alpha}\tilde{\lambda}^{\dot{\alpha}}x_{\alpha \dot{\alpha}}(y))}$ is the non-integrated vertex operator and

\begin{equation}
U(y) = [i\varepsilon^{\alpha}\tilde{\lambda}^{\dot{\alpha}}\partial x_{\alpha \dot{\alpha}}(y) -2(\tilde{\lambda}\psi)(\tilde{\lambda}\eta)]V(y)  
\label{integratedvertexop}
\end{equation}

\noindent is the integrated one. The $\lambda$'s and $\tilde{\lambda}$'s are related to the momenta
of the particles via eq. \eqref{momentumasspinors}, while $\varepsilon$ was defined in eq. \eqref{polarizations}. The amplitude is to be calculated like a superstring expectation value, with $y_1$, $y_2$ and $y_3$ fixed to any arbitrary
(different) positions over the real line, and with the following OPE's for the fields:

\begin{align}
  x_{\alpha \dot{\alpha}} (z) x_{\beta \dot{\beta}}(w) &\sim - \alpha' \epsilon_{\alpha \beta} \epsilon_{\dot{\alpha} \dot{\beta}} (\ln|z-w|+\ln|z-\bar{w}|) \nonumber \\
  \psi_{\dot{\alpha}}(z)\eta_{\dot{\beta}}(w) &\sim  \frac{\alpha' \epsilon_{\dot{\alpha}\dot{\beta}}}{z-w} \label{OPEs_gluons}
\end{align}

\indent The other OPE's are non-singular.\\
\indent The authors have shown that the above prescription reduces to the right result of eq. \eqref{mhv_gluon} in the limit $\alpha'=0$,
where open superstring theory is supposed to describe sYM. Finally, they have checked that it gives the right amplitude for
scatterings between 4 and 5 gluons including $\alpha'$ corrections, as one may compare with the results in the literature (see \cite{polchinski1} for 4 gluons).\footnote{These results have been reviewed before in \cite{gustavo}.} Stieberger and Taylor
\cite{stieberger} have also shown that it works for 6 gluons, also manipulating it to show that
the general result can be written as a linear combination of a particular kind of integrals.\\
\indent Notice that the Berkovits-Maldacena prescription makes sense only
in a 4 dimensional spacetime (we are using $SU(2)$ indices) so, in order to compare its results to usual superstrings, one has to attach the latter
to D3-branes, forcing the open superstrings to move only in 4 dimensions.\\
\indent The intriguing part of eq. \eqref{gluons_prescription} is that the action $S$ used in the averaging
is not known, i.e., the prescription looks like belonging to a string theory, except that we don't know
how to deduce it from one. We can make calculations without this information, though, only using OPE's and holomorphicity properties of
the fields.\\
\indent The fields appearing in the vertex operators are the same that appear in the superstring theory
with $\mathcal{N}$ $=2$ world-sheet supersymmetry (\cite{ooguri} and references therein). The action for this string\footnote{The fields $\bar{\eta}$ and $\bar{\psi}$ do not enter here in the context of open strings, but will appear in the next sections.} is:

\begin{equation}
 S = \frac{1}{\alpha'}\int d^2 z \left( \frac{1}{2}\partial x^{\alpha \dot{\alpha}}\bar{\partial}x_{\alpha \dot{\alpha}}+\eta^{\dot{\alpha}}\bar{\partial}\psi_{\dot{\alpha}} + \bar{\eta}^{\dot{\alpha}}\partial\bar{\psi}_{\dot{\alpha}} \right)
\label{n=2string}
\end{equation}

\indent The OPE's between these fields are the same as in Berkovits-Maldacena prescription (up to different normalizations).\\
\indent There is only one massless state in this theory, a gluon of positive helicity, i.e., this theory describes self-dual YM (\textit{without} supersymmetry!) at massless level.
This string theory lives in 4 dimensions, but with 2 of them being timelike, and not only 1 as in Minkowski spacetime.\\
\indent It would be interesting to understand better the connection between this $\mathcal{N}$ $=2$ superstring and
Berkovits-Maldacena prescription.\\

\subsection{The analogous prescription for closed strings}

\indent The purpose of this work is to write the analogue of prescription \eqref{gluons_prescription}
for ``closed strings''. Again, we won't be able to derive it from an actual
closed superstring theory (i.e. we do not know the action behind our formula), but we will show that it actually gives the right result for $N=8$ supergravity
for all $n$ ($\alpha'=0$ limit) and that it also gives the right closed superstring result for scatterings between
$n=4$ gravitons with $\alpha'$ corrections at tree level.\footnote{Again, this part was proved before in \cite{gustavo2}}\\
\indent Here is the formula:\footnote{We will use the convention $d^2z=d\sigma_1d\sigma_2$.}

\begin{align}
  \tilde{M}_n (\lambda_i, \tilde{\lambda}_i) = \frac{1}{[(\lambda_1\lambda_2)(\lambda_2\lambda_3)(\lambda_3\lambda_1)]^2} \langle &V(z_1,\bar{z}_1)V(z_2,\bar{z}_2)V(z_3,\bar{z}_3) \nonumber \\
  &\prod_{r=4}^{n}\int d^2 z_r U(z_r,\bar{z}_r) \rangle \label{THEprescription}
\end{align}

\noindent where the vertex operators are given by

\begin{equation}
 V(z,\bar{z}) \equiv \exp \left( i \lambda^{\alpha} \tilde{\lambda}^{\dot{\alpha}} x_{\alpha \dot{\alpha}}(z, \bar{z}) \right)
 \label{vvertexop}
\end{equation}

\noindent and

\begin{equation}
U(z,\bar{z}) \equiv \frac{2}{\pi \alpha'} \left[ i h^{\alpha}\tilde{\lambda}^{\dot{\alpha}}\partial x_{\alpha \dot{\alpha}}-\frac{1}{2}(\tilde{\lambda}\psi)(\tilde{\lambda}\eta) \right] \left[ i h^{\beta}\tilde{\lambda}^{\dot{\beta}}\tilde{\partial} x_{\beta \dot{\beta}}-\frac{1}{2}(\tilde{\lambda}\bar{\psi})(\tilde{\lambda}\bar{\eta})  \right] V(z,\bar{z}) 
\label{uvertexop}
\end{equation}

\indent We have used in the definition above one of the gauge parameters of eq. \eqref{gauge_graviton}
to make $\det(h_{\alpha \beta})=0$; this allows us to write $h_{\alpha \beta}= h_{\alpha}h_{\beta}$.\footnote{With
this decomposition, we will use the normalization $(h\lambda)=1$. The gauge transformation of eq. \eqref{gauge_graviton} becomes
then $\delta h_{\alpha} = c \lambda_{\alpha}$ for infinitesimal $c$. Note that, under
this transformation, the variation of the vertex operator is a total derivative.} Accordingly to
what always happens with closed strings, now the integrated vertex operators run all over the complex plane, and the non-integrated ones can be fixed to arbitrary points $z_1$, $z_2$ and $z_3$.\\
\indent The OPE's for the fields in this theory are the same as for $\mathcal{N}$ $=2$ \textit{closed} superstring theory:

\begin{align}
x_{\alpha \dot{\alpha}}(z,\bar{z})x_{\beta \dot{\beta}}(w,\bar{w}) & \sim - \alpha' \epsilon_{\alpha \beta} \epsilon_{\dot{\alpha}\dot{\beta}} \ln|z-w| \nonumber \\
\psi_{\dot{\alpha}}(z)\eta_{\dot{\beta}}(w) & \sim \frac{\alpha' \epsilon_{\dot{\alpha} \dot{\beta}}}{z-w} \label{opes}\\
\bar{\psi}_{\dot{\alpha}}(\bar{z})\bar{\eta}_{\dot{\beta}}(\bar{w}) & \sim \frac{\alpha' \epsilon_{\dot{\alpha} \dot{\beta}}}{\bar{z}-\bar{w}} \nonumber
\end{align}

\indent Again, the prescription is for a 4 dimensional spacetime, so in order to compare its results to
usual superstring theory one has to compactify 6 of the 9 spatial dimensions of the latter and
go to the limit where the sizes of this compactified dimensions are very small.\\

\subsubsection{Scattering with n=4 gravitons}

\indent Let us check the prescription \eqref{THEprescription} for the case of $n=4$ gravitons. In this case, there is only one integrated vertex operator,
inside of which only the fields $x_{\alpha \dot{\alpha}}$ will bring contributions with their OPE's. We have:

\begin{align*}
 \tilde{M}_4 = &\frac{1}{[(\lambda_1\lambda_2)(\lambda_2\lambda_3)(\lambda_3\lambda_1)]^2}\left(\frac{\alpha'}{2}\right)^2\frac{2}{\pi \alpha'} \times\\
 &\int d^2 z_4 \prod_{r<s}|z_r-z_s|^{\alpha' s_{rs}}\sum_{i=1}^3 \frac{(h_4 \lambda_i)(\tilde{\lambda}_4 \tilde{\lambda}_i)}{z_4-z_i} \sum_{j=1}^3 \frac{(h_4 \lambda_j)(\tilde{\lambda}_4 \tilde{\lambda}_j)}{\bar{z}_4-\bar{z}_j}
\end{align*}

\noindent where $s_{rs} \equiv k_r \cdot k_s = (\lambda_r\lambda_s)(\tilde{\lambda}_r\tilde{\lambda}_s)$.\\
\indent We can fix $(z_1, z_2,z_3)=(1,\infty,0)$. The integrand becomes then:

\[ |z_4 -1|^{\alpha' s_{14}} |z_4|^{\alpha' s_{34}} \left( \frac{(h_4 \lambda_1)(\tilde{\lambda}_4 \tilde{\lambda}_1)}{z_4 - 1}+\frac{(h_4 \lambda_3)(\tilde{\lambda}_4 \tilde{\lambda}_3)}{z_4} \right) \left( \frac{(h_4 \lambda_1)(\tilde{\lambda}_4 \tilde{\lambda}_1)}{\bar{z}_4 - 1}+\frac{(h_4 \lambda_3)(\tilde{\lambda}_4 \tilde{\lambda}_3)}{\bar{z}_4} \right) \]

\indent Here we have used property \eqref{conservationproperty2}.
Now we choose $(h_4)_{\alpha} = (\lambda_3)_{\alpha}/(\lambda_3 \lambda_4)$ so as to cancel the second terms inside each pair of parenthesis (note that
we are respecting the normalization $(h_4\lambda_4) = 1$).\\
\indent The amplitude becomes:

\[ \tilde{M}_4 = \frac{\alpha'}{2\pi}\frac{(\tilde{\lambda}_1\tilde{\lambda}_4)^2}{[(\lambda_1\lambda_2)(\lambda_2\lambda_3)(\lambda_3\lambda_4)]^2} \times I \]

\noindent where

\[ I = \int d^2 z_4 |z_4|^{\alpha's_{34}} |z_4-1|^{\alpha' s_{14}-2}\]

\indent Here we use the result:

\[ \int d^2 z |z|^{2a-2}|z-1|^{2b-2} = \pi \frac{\Gamma(a)\Gamma(b)\Gamma(c)}{\Gamma(a+b)\Gamma(b+c)\Gamma(c+a)} \]

\noindent with $a+b+c=1$ to write (remember that, by momentum conservation, $s_{12}=s_{34}$ and $s_{12}+s_{13}+s_{14}=0$ and so on):

\[I = \pi \frac{\Gamma(1+\frac{\alpha's_{12}}{2})\Gamma(\frac{\alpha's_{14}}{2})\Gamma(\frac{\alpha's_{13}}{2})}{\Gamma(1-\frac{\alpha's_{13}}{2})\Gamma(\frac{-\alpha's_{12}}{2})\Gamma(1-\frac{\alpha's_{14}}{2})} = - \pi \left( \frac{\alpha's_{12}}{2} \right)^2 \prod_{i=2}^4 \frac{\Gamma(\frac{\alpha's_{1i}}{2})}{\Gamma(1-\frac{\alpha's_{1i}}{2})} \]

\indent Finally, then, we have:

\begin{equation}
\tilde{M}_4 = -\frac{\alpha'^3}{8} \frac{(\tilde{\lambda}_1\tilde{\lambda}_4)^2(\tilde{\lambda}_1\tilde{\lambda}_2)^2}{(\lambda_2\lambda_3)^2(\lambda_3\lambda_4)^2} \prod_{i=2}^4 \frac{\Gamma(\frac{\alpha's_{1i}}{2})}{\Gamma(1-\frac{\alpha's_{1i}}{2})}
\label{4graviton}
\end{equation}

\indent This is the same result as one would get via usual superstring, as we show in appendix \eqref{apendiceb}.\\
\indent Remembering that the gamma function behaves like $\Gamma(x)=\frac{1}{x}+O(x^0)$ for small $x$ (also $\Gamma(1)=1$),
we see that in the limit $\alpha' = 0$ we have

\[ \tilde{M}_4(\alpha'=0) = - \frac{(\tilde{\lambda}_1\tilde{\lambda}_4)(\tilde{\lambda}_1\tilde{\lambda}_2)}{(\lambda_3\lambda_4)^2(\lambda_1\lambda_2)(\lambda_1\lambda_3)(\lambda_1\lambda_4)(\lambda_2\lambda_3)^2(\tilde{\lambda}_1\tilde{\lambda}_3)} \]

\noindent which, using property \eqref{conservationproperty} to write $(\lambda_2\lambda_3)(\tilde{\lambda}_3\tilde{\lambda}_1)+(\lambda_2\lambda_4)(\tilde{\lambda}_4\tilde{\lambda}_1)=0 $, gives the same result as eq. \eqref{n4hodges}.\\

\subsubsection{The $\alpha'$=0 limit, for arbitrary n}

\indent To evaluate this expression in the limit where $\alpha' = 0$, we will first rewrite the integrated vertex operators
using integrals over Grassmann variables:

\begin{align}
  U(z,\bar{z})=\frac{2}{\pi \alpha'}&\int d\chi d\xi \int d\tilde{\chi} d\tilde{\xi} \exp \{ i \lambda^{\alpha}\tilde{\lambda}^{\dot{\alpha}}x_{\alpha \dot{\alpha}} \nonumber \\
  &+ 2^{-1/2}\chi(\tilde{\lambda}\psi)+ 2^{-1/2}\xi(\tilde{\lambda}\eta)-i\chi \xi (h^{\alpha}\tilde{\lambda}^{\dot{\alpha}}\partial x_{\alpha \dot{\alpha}}) \nonumber \\
  &+ 2^{-1/2}\tilde{\chi}(\tilde{\lambda}\tilde{\psi})+ 2^{-1/2}\tilde{\xi}(\tilde{\lambda}\tilde{\eta})-i\tilde{\chi} \tilde{\xi} (h^{\beta}\tilde{\lambda}^{\dot{\beta}}\bar{\partial} x_{\beta \dot{\beta}}) \} \label{vertexoperator_susy}
\end{align}

\indent Evaluating all OPE's between the vertex operators (and using the other gauge parameter left for each particle to make all $(h_i h_j)=0$), the prescription \eqref{THEprescription} can be written then as

\begin{align}
  \tilde{M}_n = &\frac{1}{[(\lambda_1\lambda_2)(\lambda_2\lambda_3)(\lambda_3\lambda_1)]^2} \left( \prod_{r=4}^n \frac{2}{\pi \alpha'}\int d^2 z_r \int d\chi_r d\xi_r \int d\tilde{\chi}_r d\tilde{\xi}_r \right) \prod_{i < j} |z_i-z_j|^{\alpha's_{ij}}\times \nonumber \\
  &\exp \{-\frac{\alpha'}{2}\frac{(\tilde{\lambda}_i\tilde{\lambda}_j)}{z_i-z_j} \left( \chi_i \xi_i(h_i\lambda_j) +\chi_j\xi_j(h_j\lambda_i) + \chi_i\xi_j + \chi_j\xi_i \right)- \label{alfaprime}\\
  &-\frac{\alpha'}{2}\frac{(\tilde{\lambda}_i\tilde{\lambda}_j)}{\bar{z}_i-\bar{z}_j} \left( \tilde{\chi}_i \tilde{\xi}_i(h_i\lambda_j) +\tilde{\chi}_j\tilde{\xi}_j(h_j\lambda_i) + \tilde{\chi}_i\tilde{\xi}_j + \tilde{\chi}_j \tilde{\xi}_i \right)  \} \nonumber
\end{align}

\indent As the particles labeled 1, 2 and 3 enter the formula with simpler vertex operators, we have to remember that, for the above expression to be valid, we must set
$\chi_1=\chi_2=\chi_3=\xi_1=\xi_2=\xi_3=0$ (and the same for the ``tilded" variables).\\
\indent Note that, because of $(h_i h_j)=0$, we have (we are not performing sums over the repeated indices):

\begin{align*}
 \exp \{&-\frac{\alpha'}{2}\frac{(\tilde{\lambda}_i\tilde{\lambda}_j)}{z_i-z_j} \left( \chi_i \xi_i(h_i\lambda_j) +\chi_j\xi_j(h_j\lambda_i) + \chi_i\xi_j + \chi_j\xi_i \right)\}=\\
  &1 -\frac{\alpha'}{2}\frac{(\tilde{\lambda}_i\tilde{\lambda}_j)}{z_i-z_j} \left( \chi_i \xi_i(h_i\lambda_j) +\chi_j\xi_j(h_j\lambda_i) + \chi_i\xi_j + \chi_j\xi_i \right)
\end{align*}

\noindent because the only second order term available would be proportional to:

\begin{align*}
 &\chi_i\xi_i \chi_j\xi_j[(h_i\lambda_j)(h_j\lambda_i)-1] =\\
 &\chi_i\xi_i \chi_j\xi_j[(h_i\lambda_i)(h_j\lambda_j)+(h_ih_j)(\lambda_j\lambda_i)-1]=0
\end{align*}

\noindent with our choices for the $h$'s. The same is valid for tilded variables.\\
\indent There are many terms coming from the Berezin integrals in eq. \eqref{alfaprime}. Let us analyze them.\\
\indent First, consider a term in which the integral over, say, variable $\chi_{i_1}$ has been made when it meets a variable $\xi_{i_2}$. Then, to integrate
over $\chi_{i_2}$, one has to choose a term where it meets another $\xi$, say, $\xi_{i_3}$, and so on, and the only way to end this process is by closing a cycle: $\chi_{i_m}=\chi_{i_1}$.\\
\indent The same is valid for the tilded variables. And the same is valid for terms in which the $\chi_i\xi_i(h_i\lambda_j)$ term in one set
of parenthesis meets the $\tilde{\chi}_j\tilde{\xi}_j(h_j\lambda_i)$ term in the other parenthesis: all of them give
rise to cycles.\\
\indent Now, let us analyze the problem with these cycles.
The terms from the exponentials will be all of order $\alpha'^2$. Considering that each vertex operator (from $i=3$ on) has a factor of $\alpha'^{-1}$, the only integrals
over the complex variables $z_i$ which interest us are those who give rise to a factor of $\alpha'^{-1}$, so that the overall order will be $\alpha'^0$; the terms with this
dependence on $\alpha'$, of course, are the only ones to survive in the limit $\alpha'=0$.\\
\indent The only way to have such a factor from those integrals is like this: Grassmann variables of particle $i$ meet the variables of particle $j$ in such a way that the modulus $|z_i-z_j|^{\alpha' s_{ij}}$
has its exponent diminished by 2 (because $(\tilde{\lambda}_i\tilde{\lambda}_j)/(z_i-z_j)$ meets $(\tilde{\lambda}_i\tilde{\lambda}_j)/(\bar{z}_i-\bar{z}_j)$). To see that this is true,
in the integral over $z_i$, we separate the domain of integration between
a ``small circle'' of radius $\delta$ around $z_j$ and the ``rest''. The ``rest'' will be integrated first only up to a big radius $R$, which we may set as $\infty$ in the end:

\[ \int d^2 z_i |z_i-z_j|^{\alpha' s_{ij}-2} = \int_{\text{circle}}d^2 z_i |z_i-z_j|^{\alpha' s_{ij}-2}+\int_{\text{rest}} d^2 z_i |z_i-z_j|^{\alpha' s_{ij}-2} \]

\indent The second term is finite for $\alpha'=0$, so it must be at least of order $\alpha'^0$;
the only part that interests us is, then:

\begin{equation}
 \int_{\text{circle}}d^2 z_i |z_i-z_j|^{\alpha's_{ij}-2}=2 \pi \int_0^{\delta} d \rho \rho^{\alpha's_{ij}-1} = 2 \pi \frac{\delta^{\alpha's_{ij}}}{\alpha's_{ij}} = 2\pi \frac{1}{\alpha's_{ij}} + O(\alpha'^0)
 \label{pole}
\end{equation}

\indent This is the $\alpha'^{-1}$ factor we were looking for. The integral over $z_i$ would hence be taken care of, and, for the rest of the expression,
$z_i$ can be set equal to $z_j$ (because $\delta$ is arbitrarily small).\\
\indent So first of all: terms in which there is a mismatch between $1/(z_i-z_j)$ and $1/(\bar{z}_i-\bar{z}_j)$, giving rise to non-real integrands, cannot
produce such a $\alpha'^{-1}$ factor, and hence can be ignored.\\
\indent Second: in a cycle (beginning, as before, at particle $i_1$, going up to particle $i_m$ before it reaches $i_1$ again), we would have the following:

\begin{itemize}
 \item Integrating over $z_{i_1}$ would produce a factor of $2\pi/(\alpha' s_{i_1 i_2})$, and we would set $z_{i_1}$ everywhere else equal to $z_{i_2}$;
 \item Then we would have to integrate over $z_{i_2}$, producing a factor of $2\pi/(\alpha' s_{i_2 i_3})$ and setting $z_{i_2}$ everywhere else equal to $z_{i_3}$;
 \item And so on. Note that in the beginning of the process, the last two factors in the integrand were $|z_{i_{m-1}}-z_{i_m}|^{\alpha's_{i_{m-1}i_m}-2}$ times
 $|z_{i_{m}}-z_{i_1}|^{\alpha's_{i_{m}i_1}-2}$. But $z_{i_1}$ was set equal to $z_{i_2}$, and then to $z_{i_3}$, and so on, until it is set equal $z_{i_{m-1}}$ turning the mentioned
 integrand into $|z_{i_{m-1}}-z_{i_m}|^{\alpha'(s_{i_{m-1}i_m}+s_{i_{m}i_1})-4}$.
\end{itemize}

\begin{figure}[h]
\centering \includegraphics[width=0.5\textwidth]{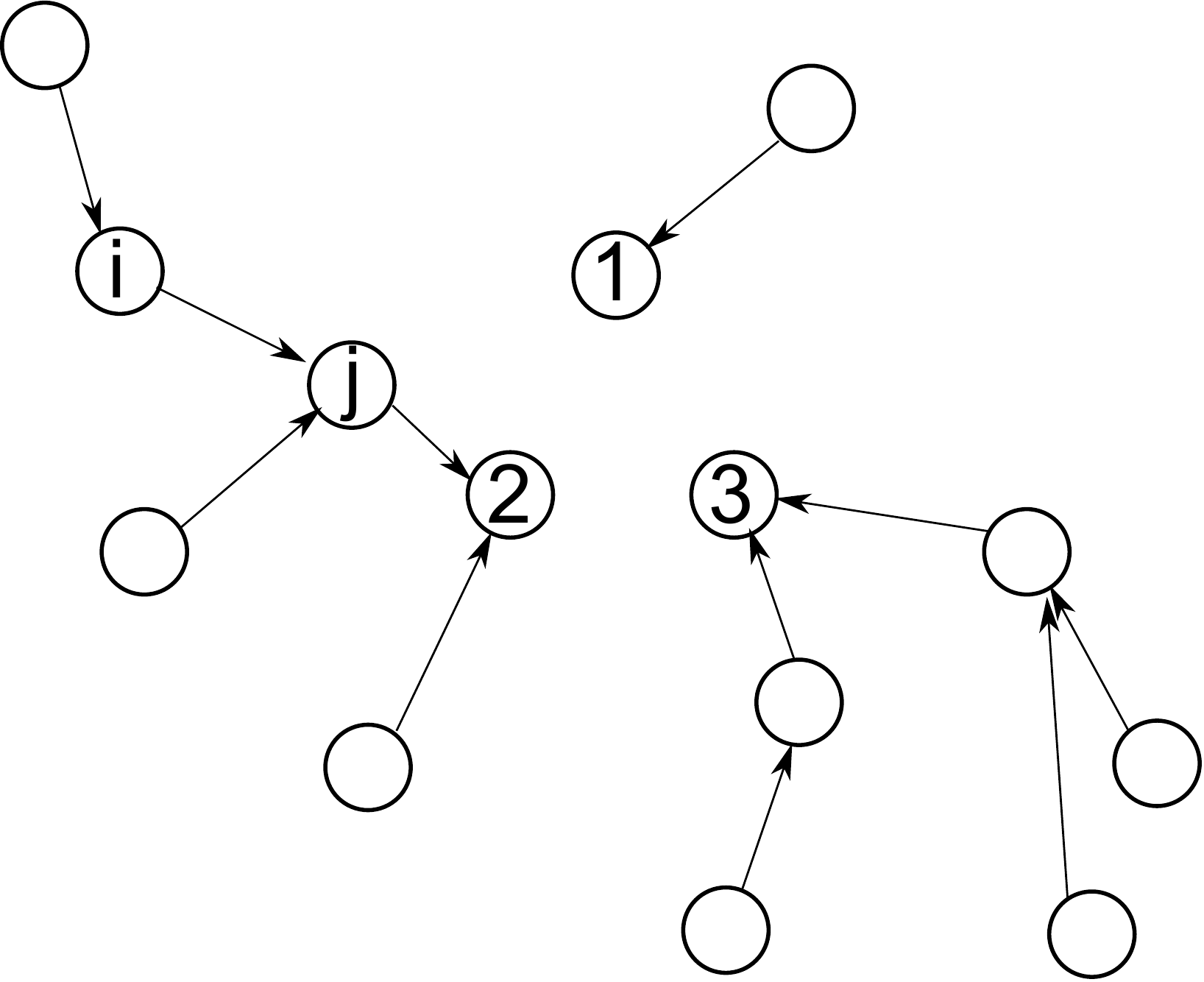}
 \caption{An example of a diagram representing a term of the formula \eqref{alfaprime}.}
 \label{diagramexample}
\end{figure}

\indent This $-4$ in the exponent is responsible for making the integral over $z_{i_m}$ (or over $z_{i_{m-1}}$, it doesn't matter) finite for $\alpha'=0$ if we follow
the same reasoning that led us to eq. \eqref{pole}.\\
\indent We conclude, then, that in a term where there is cycle,\footnote{Note that it is not important that the cycle encompasses only some or all particles.} we cannot
 cancel all the factors of $\alpha'$ in the numerator of eq. \eqref{alfaprime}.\\
\indent In short: the only terms of interest in the expression \eqref{alfaprime} are the ones in which each $\chi_i \xi_i$ portion meets its
similar $\tilde{\chi}_i \tilde{\xi}_i$, both paired with a second particle $j$, which will be paired with another particle and, to avoid chains, at some point some of these particles
will have to be paired with the ``special'' ones 1, 2 or 3. Graphically, we need to have something like figure \ref{diagramexample}.

\noindent (These diagrams are not the same as the tree diagrams proposed by Nguyen et al. \cite{nguyen} and discussed in the last chapter.)\\
\indent Each arrow in those diagrams indicates that the Berezin integral has been made over the Grassmann variables of the particle where the arrow originates,
selecting the term $\chi_i \xi_i (h_i \lambda_j)$ (if the arrow originates in $i$ and ends in $j$). Making the integral over $z_i$ gives rise
to a contribution (including all factors from eqs. \eqref{vertexoperator_susy}, \eqref{alfaprime} and \eqref{pole}):

\begin{equation}
 \frac{2}{\pi \alpha'}\frac{\alpha'^2}{4}(\tilde{\lambda}_i\tilde{\lambda}_j)^2(h_i\lambda_j)^2 \frac{2\pi}{\alpha's_{ij}} = \frac{(\tilde{\lambda}_i\tilde{\lambda}_j)}{(\lambda_i\lambda_j)}(h_i \lambda_j)^2 
 \label{matrixelement}
\end{equation}

\indent With the help of our diagrams, we will show that the expression \eqref{alfaprime} for the amplitude in the limit $\alpha'=0$ can be simplified to

\begin{equation}
\tilde{M}_n(\alpha'=0) = (-1)^{n+1}\frac{1}{[(\lambda_1\lambda_2)(\lambda_2\lambda_3)(\lambda_3\lambda_1)]^2} \phi^4_{[4}\phi^5_5...\phi^n_{n]}
\label{alfa_final}
\end{equation}

\noindent where the antisymmetrization is taken with no combinatoric factors. This is a particular form of Hodges'
formula \eqref{mhvhodges}, where the permutations $\alpha$ and $\beta$ have been taken to be the identity and, for each particle,
the arbitrary spinors $x_i$ and $y_i$ will be chosen to be $x_i=y_i=h_i$. So, by proving our diagrams agree with eq. \eqref{alfa_final}, we will show that prescription \eqref{THEprescription}
indeed reduces to $N=8$ supergravity MHV amplitudes, as it should.\\

\subsubsection*{Proof}

\indent At first glance, it seems like the expression above allows more terms than those admitted by our diagrams of the kind appearing
in figure \ref{diagramexample}. To show that both prescriptions give rise to the same analytic terms, we will also draw
the diagrams allowed by Hodges' formula. ``Our'' diagrams were drawn in solid lines, so the diagrams
of Hodges' formula will be drawn in dashed lines.\\
\indent Consider, for instance, the case $n=4$: we then only have to worry about this matrix element

\[ \phi^4_4 \equiv - \sum_{j=1}^{3} \frac{(\tilde{\lambda}_4\tilde{\lambda}_j)}{(\lambda_4\lambda_j)}(h_4 \lambda_j)^2 \]

\begin{figure}[h]
\centering \includegraphics[width=0.3\textwidth]{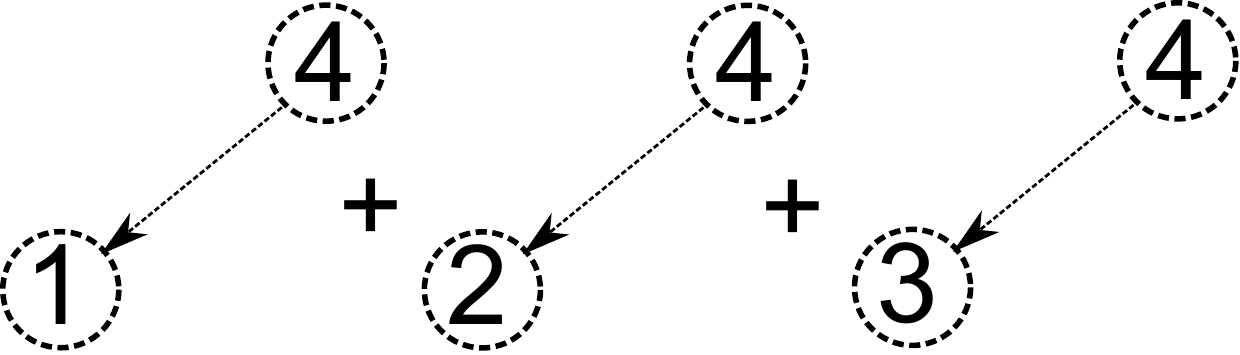}
 \caption{Hodges' diagrams for $n=4$.}
 \label{hodgesdiagramsn4}
\end{figure}

\indent This is the same as having to consider the contributions from the diagrams of figure \ref{hodgesdiagramsn4}.\\

\begin{figure}[h]
\centering \includegraphics[width=0.6\textwidth]{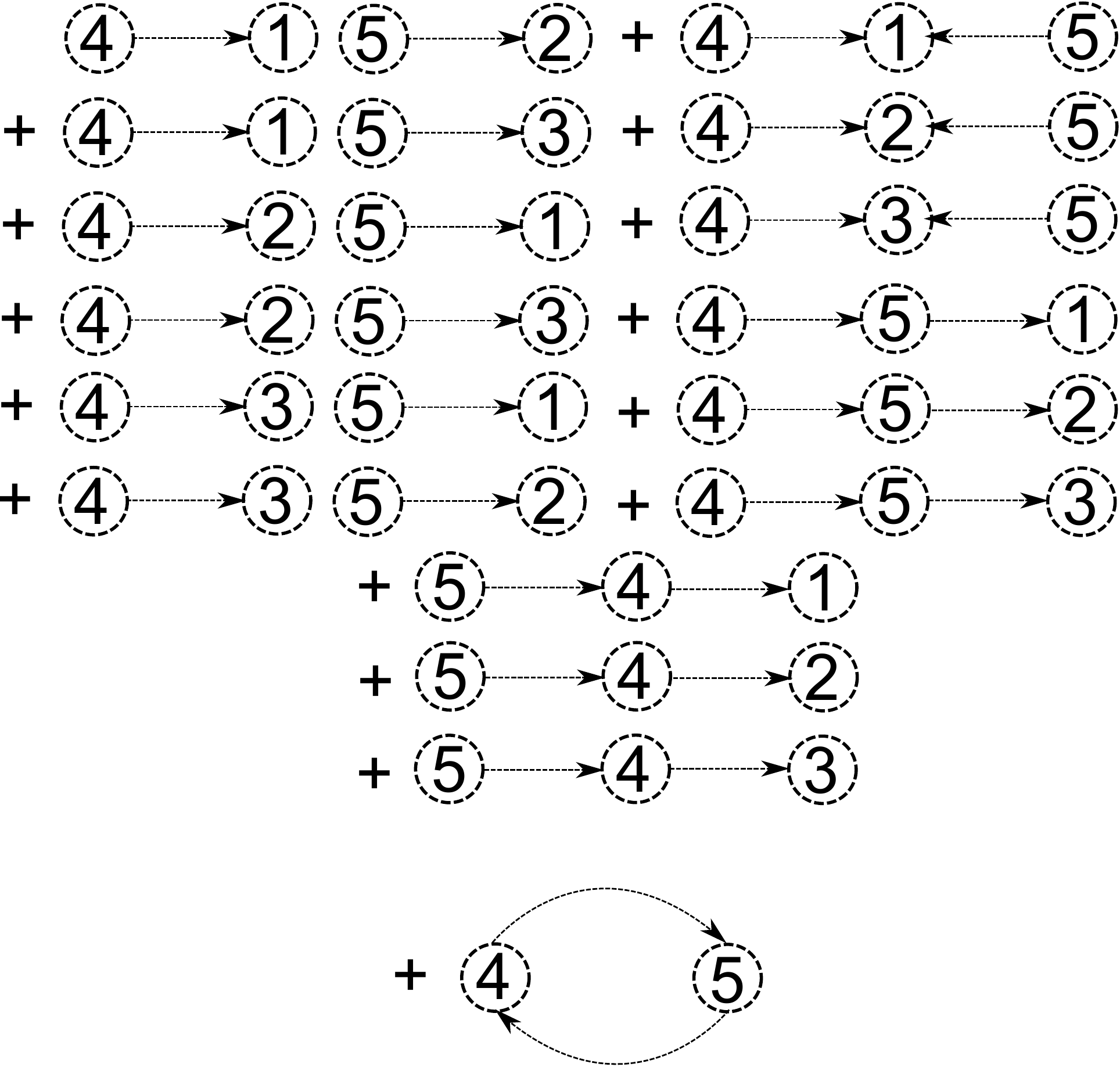}
 \caption{Some Hodges' diagrams for $n=5$.}
 \label{hodgesdiagramsn5}
\end{figure}

\noindent where the rule to each arrow is the same as before, and the terms appear with a plus sign because of one minus in the definition of $\phi_i^i$ in \eqref{hodges_definitions}
and other in eq. \eqref{alfa_final}.\\
\indent These are the same diagrams allowed by our prescription, so it gives the right result
\eqref{n4hodges} found in last chapter.\\
\indent Let us be patient and see what happens in the case $n=5$ now. The term

\[ \phi^4_{4}\phi^5_5 \]

\noindent allows the diagrams in figure \ref{hodgesdiagramsn5}.\\

\indent The cycle in this figure represent a contribution

\[ \frac{(\tilde{\lambda}_4\tilde{\lambda}_5)}{(\lambda_4\lambda_5)}(h_4 \lambda_5)^2\frac{(\tilde{\lambda}_5\tilde{\lambda}_4)}{(\lambda_5\lambda_4)}(h_5 \lambda_4)^2 \]

\indent But

\[ (h_4 \lambda_5)^2(h_5 \lambda_4)^2=[(h_4h_5)(\lambda_5\lambda_4)+(h_4\lambda_4)(h_5\lambda_5)]^2=1\]
  
\noindent making this cycle cancel the contribution from the other term:

\[ - \phi^4_{5}\phi^5_4 \]

\indent The only diagrams which give an actual contribution are also allowed by our prescription. In other words we can
say that the term above gives rise to the following diagram:

\begin{figure}[h]
\vspace{1cm}
\centering \includegraphics[width=0.3\textwidth]{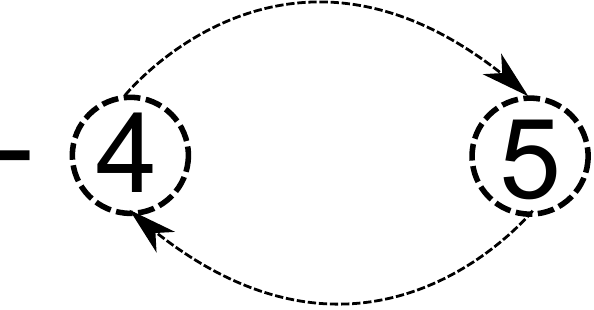}
 \caption{Another Hodges' diagrams for $n=5$.}
 \label{hodgesdiagramsn52}
\end{figure}

\begin{figure}[h]
\vspace{1cm}
\centering \includegraphics[width=0.3\textwidth]{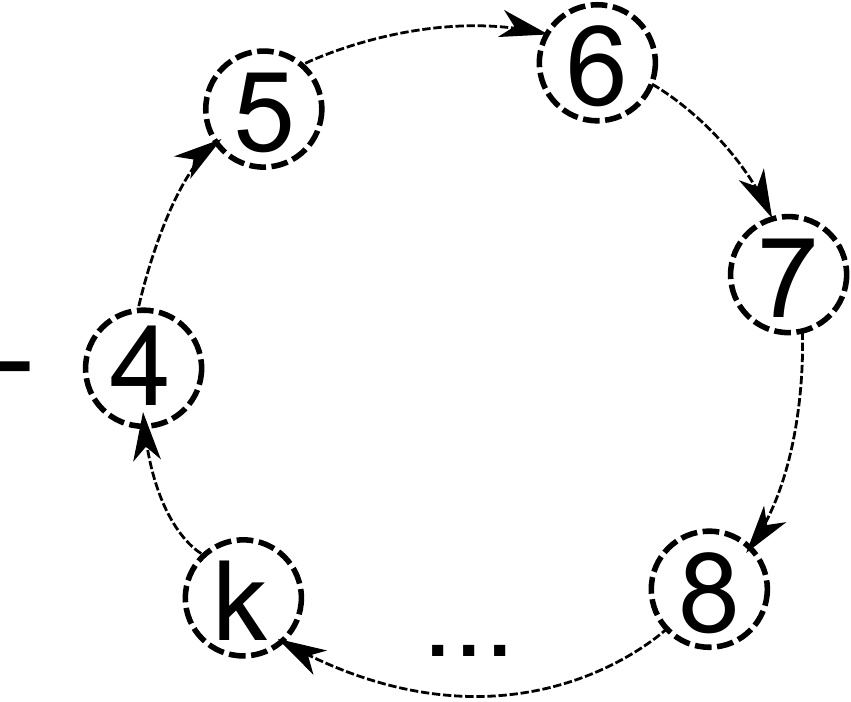}
 \caption{A cycle with ``all'' k-3 particles.}
 \label{cycleproof}
\end{figure}

\indent Note the minus sign. Then we can say that the sum over diagrams \eqref{hodgesdiagramsn5} and \eqref{hodgesdiagramsn52},
with the appropriate signs, via cancellations of cycles, leaves only diagrams allowed by our solid drawings of figure \ref{diagramexample}.\\
\indent By induction, consider now we have proved the equivalence between solid and dashed diagrams for $n=4,5,...,k-1$; let us show that it works for $n=k$.\\
\indent We saw that the first term in the sum,

\[ \phi^4_{4}\phi^5_5...\phi^k_{k} \text{ ,}\]

\noindent corresponds to all allowed solid diagrams, but also to diagrams with cycles, all of them with
a + sign ($(-1)^{k+1}$ from the overall factor in eq. \eqref{alfa_final} and other $(-1)^{k-3}$ because of the - sign in the definition of $\phi_i^i$).\\
\indent But, as we saw in the case $n=5$, terms with two indices interchanged and the others untouched, like

\[ -\phi^4_{5}\phi^5_4...\phi^k_{k} \text{ ,} \]

\noindent correspond to diagrams with 2 particles closed in a cycle, with a - sign in front,\footnote{In the case $n=5$, there
were only particles $4$ and $5$ in the diagram \ref{hodgesdiagramsn52}, but here we allow the other particles
involved in the process to attach to the cycle. This does not modify our argument, because
all ways of attaching particles to, say, the $(4,5)$ cycle are allowed by both terms above, with different signs,
hence canceling each other.} hence canceling these diagrams in the sum with the previous ones.\\
\indent Similarly, the set of permutations of three determined indices (like 4, 5 and 6, leaving the other indices untouched) has the property of canceling terms
with cycles involving 3 particles, which are not allowed by our (solid) prescription either.\\
\indent We can keep doing this, and the only terms which are left after using everything assumed by the induction hypothesis are the ones
with permutations which move \textit{all} indices.\footnote{Permutations where all indices move but only inside of two or more disjoint sets are also included in the previous smaller $n$ cases.} One of these terms is

\[ (-1)^{k} \phi^4_{5}\phi^5_6...\phi^k_{4} \text{ ,}\]

\noindent which corresponds to figure \eqref{cycleproof}

\noindent with an overall - sign (because of the overall $(-1)^{k+1}$ times the $(-1)^{k-4}$ coming from the permutations);
the other permutations left differ by the one above by a even number of transpositions, corresponding then
to other cycles involving all particles and also appearing with the - sign. These diagrams
were the last ones we needed to get rid of.\\
\indent We see, then, that our prescription \eqref{THEprescription}, in the limit $\alpha'=0$, gives rise to terms we can graphically
represent by the solid diagrams of figure \ref{diagramexample}, which are the same terms allowed by eq. \eqref{alfa_final}.\\
\indent This finishes our proof.\\

\subsubsection*{A by-product}

\indent As a by-product of our proof, we have also found a new method of calculation of MHV amplitudes for gravitons, similar
to tree diagrams proposed by Nguyen et al. \cite{nguyen}. We review the method for the sake of completeness:

\begin{itemize}
 \item In an amplitude for $n$ gravitons, take $3$ to be treated specially;
 \item Draw vertices for the particles and connect them with arrows, making all possible tree diagrams with these rules:
   \subitem All $n-3$ ``non-special'' particles must enter;
   \subitem At least one of the special particles $1$, $2$ and $3$ must enter;
   \subitem Each non-special vertex must be the origin of one and only one arrow, but they can be
   the endpoints of arbitrarily many arrows;
   \subitem Special vertices can be only endpoints.
  \item The diagrams thus drawn are like the ones in figure \ref{diagramexample}: tree diagrams
  where there are ``flows'' ending in vertices $1$, $2$ and $3$;
  \item For an arrow originating in vertex $i$ and ending in vertex $j$, put a term as defined in eq. \eqref{matrixelement};
  \item Put an overall factor $\frac{1}{[(\lambda_1\lambda_2)(\lambda_2\lambda_3)(\lambda_3\lambda_1)]^2}$ in the end.
\end{itemize}

\indent This method still makes reference to the $h$'s, upon which we have imposed
the constraints $(h_ih_j)=0$, being hence less direct than the method of \cite{nguyen}.


\section{Conclusions}

\indent Berkovits and Maldacena \cite{berkovits} have shown that MHV amplitudes also simplify
in the context of open superstring theory, proposing a prescription (formula \eqref{gluons_prescription}) for them
which is simpler than using usual superstrings, while giving the appropriate results, at least
for the verified cases of 4, 5 and 6 gluons (see \cite{berkovits} and \cite{stieberger}), and also reducing to the correct sYM result in the
$\alpha'=0$ limit.\\
\indent We have shown here that that prescription, naturally extended to closed superstrings,
also gives the correct results for the MHV amplitudes of gravitons, at least for the case of
4 gravitons and for arbitrary number of gravitons in the limit $\alpha'=0$. To this last result,
it was very important to rely upon the formula recently found by Hodges \cite{hodges}.\\
\indent Some questions remain opened, as the possible connections between
the prescription \eqref{THEprescription} and the closed superstring with $\mathcal{N}$ $=2$
world-sheet supersymmetry. Less clear is the connection with usual superstring theory
(with only $\mathcal{N}$ $=1$ supersymmetry), because, although both seem to
give the same results, including $\alpha'$ corrections, it is not known how to derive
the Berkovits-Maldacena formula from a string action.\\
\indent Results for amplitudes of gluons \cite{witten} and gravitons \cite{cachazo,cachazo2} have also been extended
to non-MHV configurations. It would also be interesting, hence, to extend the prescription
used in this work to more general configurations.\\


\subsubsection*{Acknowledgments}
I would like to thank my supervisor Nathan Berkovits, who has suggested me this
subject for my master thesis and helped me throughout the development of the work. I also thank professors
Horatiu Nastase and Diego Trancanelli (the examiners of my thesis) for very useful tips and corrections. I thank finally
FAPESP for financial support via the grant 2011/02699-6 and IFT-Unesp for providing
good teaching and infrastructure for their students.

\appendix
\section{Comparing our MHV scattering amplitude for n=4 gravitons with usual superstring theory}
\label{apendiceb}

\noindent One can find in the literature the result for the scattering amplitude between
4 gravitons (with arbitrary polarizations) in superstring theory. \cite{polchinski1} gives it like this:

\begin{equation}
 \tilde{M}_4= -\frac{i\kappa^2\alpha'^3}{4}\frac{\Gamma(-\frac{1}{4}\alpha's)\Gamma(-\frac{1}{4}\alpha't)\Gamma(-\frac{1}{4}\alpha'u)}{\Gamma(1+\frac{1}{4}\alpha's)\Gamma(1+\frac{1}{4}\alpha't)\Gamma(1+\frac{1}{4}\alpha'u)}K(h_1,h_2,h_3,h_4)
 \label{resultado}
\end{equation}

\indent Before writing down the kinematic factor $K$, let us simplify things by writing each $h$ as $h_{\mu\nu}=a_{\mu}a_{\nu}$. We will have
$a_{\mu} \rightarrow a_{\alpha \dot{\alpha}}=h_{\alpha}\tilde{\lambda}_{\dot{\alpha}}$ for the positive helicity gravitons\footnote{According to eq. \eqref{polarization_graviton} and using one gauge parameter to
make $h_{\alpha \beta}=h_{\alpha}h_{\beta}$.} and $a_{\mu} \rightarrow a_{\alpha \dot{\alpha}}=\lambda_{\alpha}\tilde{h}_{\dot{\alpha}}$ for the negative helicity.\footnote{According to eq. \eqref{polarization_graviton2} and using one gauge parameter to
make $\tilde{h}_{\dot{\alpha} \dot{\beta}}=\tilde{h}_{\dot{\alpha}}h_{\dot{\beta}}$.} With these
conventions, the factor $K$ simplifies to a square:

\begin{equation}
K(h_1,h_2,h_3,h_4) = \left[ K'(a_1,a_2,a_3,a_4) \right]^2
 \label{factorK2}
\end{equation}

\noindent where the new $K'$ is given by:

\begin{equation}
  K'(a_1,a_2,a_3,a_4) = \frac{1}{8}\left[ 4 C^{\mu}_{1\nu}C^{\nu}_{2\rho}C^{\rho}_{3\sigma}C^{\sigma}_{4\mu}-C^{\mu}_{1\nu}C^{\nu}_{2\mu}C^{\rho}_{3\sigma}C^{\sigma}_{4\rho} \right]+(1324)+(1423)
 \label{factorK'}
\end{equation}

\noindent with

\begin{equation}
 C^{\mu}_{i\nu} \equiv k_i^{\mu}a_{i\nu}-k_{i\nu}a_i^{\mu}
 \label{Cdef}
\end{equation}

\noindent and the parenthesis indicate other permutations. This $K'$ is exactly what appears
in the analogous amplitude for 4 gluons with polarizations $a_i$.\\
\indent The ratio between the gamma functions in eq. \eqref{resultado} agrees with our result \eqref{4graviton}, because the Mandelstam
variables are:

\begin{align}
 &s \equiv - (k_1+k_2)^2=-2(k_1\cdot k_2)=-2s_{12}\nonumber\\
 &t \equiv - (k_1+k_3)^2=-2(k_1\cdot k_3)=-2s_{13}\\
 &u \equiv - (k_1+k_4)^2=-2(k_1\cdot k_4)=-2s_{14}\nonumber
\end{align}

\indent We then only need to worry about showing that the factor $K$ reduces
to what appears in our result \eqref{4graviton} when we specialize to an MHV amplitude.\\
\indent We will choose gravitons 1 and 2 to have negative helicity. Half of the terms in eq. \eqref{factorK'} are:

\begin{align}
  &C^{\mu}_{1\nu}C^{\nu}_{2\mu}C^{\rho}_{3\sigma}C^{\sigma}_{4\rho} =\nonumber\\
  &=4[(a_1\cdot a_2)(k_1\cdot k_2)-(a_1\cdot k_2)(a_2\cdot k_1)][(a_3\cdot a_4)(k_3\cdot k_4)-(a_3\cdot k_4)(a_4\cdot k_3)]\label{varios1}\\
  &C^{\mu}_{1\nu}C^{\nu}_{3\mu}C^{\rho}_{2\sigma}C^{\sigma}_{4\rho} =\nonumber\\
  &=4[(a_1\cdot a_3)(k_1\cdot k_3)-(a_1\cdot k_3)(a_3\cdot k_1)][(a_2\cdot a_4)(k_2\cdot k_4)-(a_2\cdot k_4)(a_4\cdot k_2)] \label{varios2}\\
  &C^{\mu}_{1\nu}C^{\nu}_{4\mu}C^{\rho}_{2\sigma}C^{\sigma}_{3\rho} =\nonumber\\
  &=4[(a_1\cdot a_4)(k_1\cdot k_4)-(a_1\cdot k_4)(a_4\cdot k_1)][(a_2\cdot a_3)(k_2\cdot k_3)-(a_2\cdot k_3)(a_3\cdot k_2)] \label{varios3}
\end{align}

\indent But $(a_3\cdot a_2) = (h_3\lambda_2)(\tilde{\lambda}_3\tilde{h}_2)$ and $(a_3\cdot k_2) = (h_3 \lambda_2)(\tilde{\lambda}_3\tilde{\lambda}_2)$. Hence, choosing
$(h_3)^{\alpha}=(\lambda_2)^{\alpha}/(\lambda_2\lambda_3)$ the part \eqref{varios3} vanishes. The part
\eqref{varios2} will similarly vanish for $(h_4)^{\alpha}=(\lambda_2)^{\alpha}/(\lambda_2\lambda_4)$, making
also $(a_3\cdot a_4)=0$\\
\indent Finally, for \eqref{varios1}, we can make $(a_1\cdot k_2)$ vanish by making $(\tilde{h}_1)^{\dot{\alpha}}=(\tilde{\lambda}_2)^{\dot{\alpha}}/(\tilde{\lambda}_2\tilde{\lambda}_1)$. Making
yet $(\tilde{h}_2)^{\dot{\alpha}}=(\tilde{\lambda}_1)^{\dot{\alpha}}/(\tilde{\lambda}_1\tilde{\lambda}_2)$ we have this contribution to $K'$ (mind the $1/8$ factor in eq. \eqref{factorK'}):

\begin{align}
&\frac{4}{8}(a_1\cdot a_2)(k_1\cdot k_2)(a_3\cdot k_4)(a_4\cdot k_3)\\
&= \frac{1}{2}(\lambda_1\lambda_2)^2\frac{(\lambda_2\lambda_4)(\tilde{\lambda}_3\tilde{\lambda}_4)}{(\lambda_2\lambda_3)}\frac{(\lambda_2\lambda_3)(\tilde{\lambda}_4\tilde{\lambda}_3)}{(\lambda_2\lambda_4)} \nonumber
\end{align}

\indent We can just cancel $(\lambda_2\lambda_3)$ and $(\lambda_2\lambda_4)$; or we can cancel the term $(\lambda_2\lambda_4)$ in the numerator and the denominator, use property \eqref{conservationproperty}
to write $(\lambda_2\lambda_3)(\tilde{\lambda}_4\tilde{\lambda}_3)=-(\lambda_2\lambda_1)(\tilde{\lambda}_4\tilde{\lambda}_1)$ and finally use $s_{12}=s_{34}$ to write $(\tilde{\lambda}_3\tilde{\lambda}_4)=(\tilde{\lambda}_1\tilde{\lambda}_2)(\lambda_1\lambda_2)/(\lambda_3\lambda_4)$. We arrive thus at two forms for this contribution:

\begin{equation}
-\frac{1}{2}(\lambda_1\lambda_2)^2(\tilde{\lambda}_3\tilde{\lambda_4})^2= -\frac{1}{2}(\lambda_1\lambda_2)^4\frac{(\tilde{\lambda}_1\tilde{\lambda}_2)(\tilde{\lambda}_1\tilde{\lambda}_4)}{(\lambda_2\lambda_3)(\lambda_3\lambda_4)} 
 \label{onepartofK'}
\end{equation}

\indent But $K'$ (eq. \eqref{factorK'}) also has contributions coming from the terms where all $C_i$'s are coupled. With the conventions we adopted for the polarizations above, these contributions reduce to:

\begin{align}
   C^{\mu}_{1\nu}C^{\nu}_{2\rho}C^{\rho}_{3\sigma}C^{\sigma}_{4\mu}+C^{\mu}_{1\nu}C^{\nu}_{4\rho}C^{\rho}_{2\sigma}C^{\sigma}_{3\mu}=&-2(a_1\cdot a_4)(k_1\cdot k_2)(a_2\cdot k_3)(a_3\cdot k_4) -\label{a10}\\
   &-2(a_1\cdot a_2)(k_1\cdot a_4)(k_2\cdot k_3)(a_3\cdot k_4)\nonumber\\
   C^{\mu}_{1\nu}C^{\nu}_{3\rho}C^{\rho}_{2\sigma}C^{\sigma}_{4\mu}=&-(a_1\cdot a_4)(k_1\cdot a_3)(k_3\cdot a_2)(k_2\cdot k_4)+ \label{a11}\\
   &+(a_1\cdot a_4)(k_1\cdot a_3)(k_3\cdot k_2)(a_2\cdot k_4)+ \nonumber\\
   &+(a_1\cdot a_3)(k_1\cdot a_4)(k_3\cdot a_2)(k_2\cdot k_4)-\nonumber \\
   &-(a_1\cdot a_3)(k_1\cdot a_4)(k_3\cdot k_2)(a_2\cdot k_4)\nonumber
\end{align}

\indent Contracting $k_1+k_2+k_3+k_4=0$ with $a_2$ to write $(a_2\cdot k_4)=-(a_2\cdot k_3)$ makes the first and second lines of \eqref{a11} join; we can similarly
make the third and fourth lines join, simplifying eq. \eqref{a11} to:

\begin{align}
 C^{\mu}_{1\nu}C^{\nu}_{3\rho}C^{\rho}_{2\sigma}C^{\sigma}_{4\mu}=&+(a_1\cdot a_4)(k_1\cdot a_3)(k_3\cdot a_2)(k_2\cdot k_1)- \label{a12}\\
   &-(a_1\cdot a_3)(k_1\cdot a_4)(k_3\cdot a_2)(k_2\cdot k_1)\nonumber
\end{align}

\indent Now the first terms of \eqref{a10} and \eqref{a12} are the same (because $(a_3\cdot k_4)=-(a_3\cdot k_1)$), so they sum. Less trivially, the second terms in both equations are
also the same (it is easier to write down the products in terms of spinors to see so). Hence, apart from eq. \eqref{onepartofK'}, $K'$ also has this contribution (putting all numeric factors of eqs. \eqref{factorK'}, \eqref{a10} and \eqref{a12}):

\begin{align}
 &\frac{3}{2}(a_1\cdot a_4)(k_1\cdot a_3)(k_3\cdot a_2)(k_2\cdot k_1)- \label{a13}\\
   &-\frac{3}{2}(a_1\cdot a_3)(k_1\cdot a_4)(k_3\cdot a_2)(k_2\cdot k_1)\nonumber
\end{align}

\indent But:

\begin{align}
 &(a_1\cdot a_4)(k_1\cdot a_3)-(a_1\cdot a_3)(k_1\cdot a_4)=\\
 &=\frac{(\lambda_1\lambda_2)(\tilde{\lambda}_2\tilde{\lambda}_4)}{(\lambda_2\lambda_4)(\tilde{\lambda}_2\tilde{\lambda}_1)}\frac{(\lambda_1\lambda_2)(\tilde{\lambda}_1\tilde{\lambda}_3)}{(\lambda_2\lambda_3)}-\frac{(\lambda_1\lambda_2)(\tilde{\lambda}_2\tilde{\lambda}_3)}{(\lambda_2\lambda_3)(\tilde{\lambda}_2\tilde{\lambda}_1)}\frac{(\lambda_1\lambda_2)(\tilde{\lambda}_1\tilde{\lambda}_4)}{(\lambda_2\lambda_4)}=\nonumber\\
 &=\frac{(\lambda_1\lambda_2)^2}{(\tilde{\lambda}_2\tilde{\lambda}_1)(\lambda_2\lambda_3)(\lambda_2\lambda_4)}\left[ (\tilde{\lambda}_2\tilde{\lambda}_4)(\tilde{\lambda}_1\tilde{\lambda}_3) -(\tilde{\lambda}_2\tilde{\lambda}_3)(\tilde{\lambda}_1\tilde{\lambda}_4)\right] =\nonumber\\
 &=\frac{(\lambda_1\lambda_2)^2}{(\tilde{\lambda}_2\tilde{\lambda}_1)(\lambda_2\lambda_3)(\lambda_2\lambda_4)}(\tilde{\lambda}_1\tilde{\lambda}_2)(\tilde{\lambda}_3\tilde{\lambda}_4) \nonumber\\
 &=-\frac{(\lambda_1\lambda_2)^2 (\tilde{\lambda}_3\tilde{\lambda}_4)}{(\lambda_2\lambda_3)(\lambda_2\lambda_4)}\nonumber
\end{align}

\indent Putting this in eq. \eqref{a13}, we can rewrite this contribution to $K'$ as:

\begin{align}
&-\frac{3}{2}\frac{(\lambda_1\lambda_2)^2 (\tilde{\lambda}_3\tilde{\lambda}_4)}{(\lambda_2\lambda_3)(\lambda_2\lambda_4)}\frac{(\lambda_2\lambda_3)(\tilde{\lambda}_1\tilde{\lambda}_3)}{(\tilde{\lambda}_1\tilde{\lambda}_2)}(\lambda_1\lambda_2)(\tilde{\lambda}_1\tilde{\lambda}_2)= \label{otherpartofK'}\\
&= \frac{3}{2}(\lambda_1\lambda_2)^2 (\tilde{\lambda}_3\tilde{\lambda}_4)^2\nonumber
\end{align}

\indent This sums with eq. \eqref{onepartofK'} to give

\begin{equation}
   K' = (\lambda_1\lambda_2)^2(\tilde{\lambda}_3\tilde{\lambda_4})^2= (\lambda_1\lambda_2)^4\frac{(\tilde{\lambda}_1\tilde{\lambda}_2)(\tilde{\lambda}_1\tilde{\lambda}_4)}{(\lambda_2\lambda_3)(\lambda_3\lambda_4)} 
 \label{k'}
\end{equation}

\indent We wrote this in the second, more complicated form, because like this we see that $K'$ squared (i.e. the $K$ of eq. \eqref{factorK2}) gives in eq. \eqref{resultado} the same
factor in front of the $\Gamma$'s than in our result \eqref{4graviton},\footnote{Up to multiplicative constants.} times
the $(\lambda_1\lambda_2)^8$ we factored out in eq. \eqref{factorgravitons}



\newpage

\begin{thebibliography}{999}

\bibitem{berkovits}
N. Berkovits and J. Maldacena, {\it Fermionic T-Duality, Dual Superconformal
Symmetry, and the Amplitude/Wilson Loop Connection}, \href{http://arxiv.org/abs/0807.3196}{arXiv:0807.3196} [hep-th]
\bibitem{parke}
S. Parke and T. Taylor, {\it An Amplitude For N Gluon Scattering}, Phys. Rev. Lett.
{\bf 56}, 2459 (1986)
\bibitem{berends}
F. A. Berends and W. T. Giele, {\it Recursive Calculations For Processes With N
Gluons}, Nucl. Phys. B {\bf 306}, 759 (1988)
\bibitem{nair}
V. Nair, {\it A Current Algebra For Some Gauge Theory Amplitudes}, Phys. Lett. {\bf B214} 215 (1988)
\bibitem{stieberger}
S. Stieberger and T.R. Taylor, {\it Maximally Helicity Violating Disk Amplitudes, Twistors and Transcendental Integrals},
\href{http://arxiv.org/abs/1204.3848}{arXiv:1204.3848} [hep-th]
\bibitem{gustavo2} Unpublished work by G.M. Monteiro.
\bibitem{berends2}
F.A. Berends, W.T. Giele and H. Kuijf, {\it On relations between multi-gluon and multi-graviton scattering},
Physics Letters B {\bf 211}, 91-94 (1988)
\bibitem{witten0}
E. Witten, {\it Perturbative Gauge Theory As A String Theory In Twistor Space},
\href{http://arxiv.org/abs/hep-th/0312171}{arXiv:hep-th/0312171}
\bibitem{tranca}
S. Giombi, R. Ricci, D. Robles-Llana and D. Trancanelli, {\it A Note on Twistor Gravity Amplitudes},
\href{http://arxiv.org/abs/hep-th/0405086}{arXiv:hep-th/0405086}
\bibitem{bedford}
J. Bedford, A. Brandhuber, B. J. Spence and G. Travaglini, {\it A Recursion Relation for Gravity Amplitudes},
Nucl. Phys. B {\bf 721}, 98 (2005) \href{http://arxiv.org/abs/hep-th/0502146}{arXiv:hep-th/0502146v2}
\bibitem{elvang}
H. Elvang and D. Z. Freedman, {\it Note on graviton MHV amplitudes}, JHEP 0805, 096 (2008) \href{http://arxiv.org/abs/0710.1270}{arXiv:0710.1270} [hep-th]
\bibitem{mason}
L. Mason and D. Skinner, {\it Gravity, Twistors and the MHV Formalism} \href{http://arxiv.org/abs/0808.3907}{arXiv:0808.3907v2} [hep-th]
\bibitem{nguyen}
D. Nguyen, M. Spradlin, A. Volovich and C. Wen, {\it The Tree Formula for MHV Graviton Amplitudes},
\href{http://arxiv.org/abs/0907.2276}{arXiv:0907.2276v2} [hep-th]
\bibitem{hodges2}
A. Hodges, {\it New expressions for gravitational scattering
amplitudes}, \href{http://arxiv.org/abs/1108.2227}{arXiv:1108.2227v2} [hep-th]
\bibitem{hodges}
A. Hodges, {\it A simple formula for gravitational MHV
amplitudes}, \href{http://arxiv.org/abs/1204.1930}{arXiv:1204.1930v1} [hep-th]
For a combined version of the two articles above, see A. Hodges, {\it New expressions for gravitational scattering amplitudes}, JHEP, DOI 10.1007/JHEP07(2013)075
\bibitem{ooguri}
H. Ooguri and C. Vafa, {\it Selfduality and N=2 String Magic}, Mod. Phys. Lett. A {\bf 5} (1990)
1389;\\
H. Ooguri and C. Vafa, {\it Geometry of N=2 Strings}, Nucl. Phys. B {\bf 361}, 469 (1991);\\
H. Ooguri and C. Vafa, {\it N=2 Heterotic Strings}, Nucl. Phys. B {\bf 367}, 83 (1991).
\bibitem{witten}
F. Cachazo, P. Svrcek and E. Witten, {\it MHV Vertices And Tree Amplitudes In Gauge Theory},
\href{http://arxiv.org/abs/hep-th/0403047}{arXiv:hep-th/0403047}
\bibitem{cachazo}
F. Cachazo and D. Skinner, {\it Gravity from Rational Curves},
\href{http://arxiv.org/abs/1207.0741}{arXiv:1207.0741v1} [hep-th]
\bibitem{cachazo2}
F. Cachazo, L. Mason and D. Skinner, {\it Gravity in Twistor Space and its Grassmannian Formulation},
\href{http://arxiv.org/abs/1207.4712}{arXiv:1207.4712v1} [hep-th]
\bibitem{gustavo}
G.M. Monteiro, {\it MHV Tree Amplitudes in Super-Yang-Mills and
in Superstring Theory}, master dissertation under supervision of N. Berkovits, IFT-Unesp (2010),
Link: \url{http://www.athena.biblioteca.unesp.br/exlibris/bd/bft/33015015001P7/2010/monteiro_gm_me_ift.pdf}
\bibitem{klt}
H. Kawai, D. C. Lewellen and S. H. H. Tye, {\it A relation between tree amplitudes of closed and open strings},
Nucl. Phys. B {\bf 269}, 1 (1986).
\bibitem{polchinski1}
J. Polchinski, {\it String Theory, 2vol}, Cambridge University Press,
Cambridge (1998).

\end{thebibliography}
\end{document}